\documentclass[%
 preprint,
superscriptaddress,
 amsmath,amssymb,
 aps,
]{revtex4-1}

\usepackage{graphicx}% Include figure files
\usepackage{dcolumn}% Align table columns on decimal point
\usepackage{bm}% bold math
\usepackage[caption=false]{subfig}
\usepackage{color}
\newcommand{\red}[1]{\textcolor{black}{#1}}
\usepackage[capitalize]{cleveref}

\begin{document}

\preprint{APS/123-QED}

\title{Frictional boundary layer effect on\\ vortex condensation in rotating turbulent convection}
%\title{Can Vortex Condensation Survive\\ Frictional Boundary Layers in Rotating Turbulent Convection?}
\author{Andr\'es J. \surname{Aguirre Guzm\'an}}
\author{Matteo Madonia}
\affiliation{Fluids and Flows group, Department of Applied Physics and J. M. Burgers Centre for Fluid Dynamics, Eindhoven University of Technology, P.O. Box 513, 5600 MB Eindhoven, The Netherlands}
\author{Jonathan S. Cheng}
\affiliation{Department of Mechanical Engineering, University of Rochester, Rochester, NY 14627, USA}
%\email{a.j.aguirre.guzman@tue.nl}
\author{Rodolfo \surname{Ostilla-M\'onico}}
\affiliation{Department of Mechanical Engineering, University of Houston, Houston, TX 77004, USA.}
\author{Herman J. H. Clercx}
\author{Rudie P. J. Kunnen}
\affiliation{Fluids and Flows group, Department of Applied Physics and J. M. Burgers Centre for Fluid Dynamics, Eindhoven University of Technology, P.O. Box 513, 5600 MB Eindhoven, The Netherlands}

\date{\today}

\begin{abstract}
We \red{perform direct numerical simulations of} rotating Rayleigh\red{--}B\red{\'e}nard convection of fluids with low ($Pr=0.1$) and high ($Pr=5$) Prandtl numbers in a horizontally periodic layer with no-slip top and bottom boundaries. At both Prandtl numbers, we demonstrate the presence of an upscale transfer of kinetic energy that leads to the development of domain-filling vortical structures. Sufficiently strong buoyant forcing and rotation foster the quasi-two-dimensional turbulent state of the flow, despite the \red{formation} of plume-like vertical \red{disturbances promoted} by \red{so-called} Ekman pumping from the viscous boundary layer.
\end{abstract}

\maketitle

% Motivation for field of study and broad problem.

While turbulent flows in nature are intrinsically three-dimensional (3D), many geophysical and astrophysical flows exhibit quasi-two-dimensional (Q2D) behavior \cite{charney1971geostrophic,pedlosky2013geophysical}. The partial reduction of dimensionality, which may be imposed by rapid rotation \red{\cite{smith1999transfer, campagne2014direct, godeferd2015structure,seshasayanan2018condensates}}, intense magnetic fields \red{\cite{alexakis2011two, xia2017two}} or by \red{geometrical confinement \cite{kellay1995experiments,smith1996crossover,musacchio2019condensate}}, offers an intermediate state between the classical problems of 3D and 2D turbulence. In pure 3D turbulent flows, \red{energy cascades downscale} according to the well-established Kolmogorov theory \red{\cite{kolmogorov1941local,frisch1995}}. \red{Conversely, 2D turbulent flows exhibit upscale energy transfer} in accordance with the theory of Kraichnan and Batchelor \cite{kraichnan1967inertial,batchelor1969computation}. \red{The upscale transfer leads to spectral energy condensation at the largest scales and the formation of long-lived coherent structures \cite{smith1994finite, chertkov2007dynamics, frishman2018turbulence, boffetta2012two}, where upscale transfer is eventually balanced by frictional effects in a finite-size domain or by imposing} large-scale frictional damping \cite{van2019condensates}\red{.} In Q2D turbulence, the upscale transfer of kinetic energy occurs despite the introduction of 3D perturbations \red{\cite{vallis1993generation, alexakis2018cascades}}. Spectral condensation can manifest as zonal flows and large-scale vortices in the atmosphere \red{and interior} of planets and in the oceans \cite{galperin2004ubiquitous, aurnou2015rotating, vallis2017atmospheric}.

% Motivation for particular facet of the problem.

Turbulence in oceanic, atmospheric and planetary settings is often driven by buoyancy and further shaped by rotation. \red{A simple model for these flows is rotating Rayleigh--B\'enard convection (RRBC), the flow in a layer of fluid confined between two parallel horizontal plates, heated from below and cooled from above, and subject to rotation about a vertical axis.}  Recent studies \red{in RRBC} in a horizontally periodic box with stress-free (SF) \red{boundary conditions (BCs) on} top and bottom \red{plates have identified} upscale energy transfer and accompanying large-scale vortices (LSVs) \cite{julien2012statistical,rubio2014upscale,favier2014inverse,guervilly2014large,stellmach2014approaching,kunnen2016transition}. This Neumann-type boundary condition provides the most favorable circumstances for the establishment of horizontal flows. No-slip (NS) \red{BCs}, on the other hand, promote the development of viscous Ekman boundary layers that actively enhance vertical motions through so-called Ekman pumping \cite{pedlosky2013geophysical}, leading to \red{bulk perturbations of small horizontal scale~\cite{stellmach2014approaching,kunnen2016transition} that} decrease the degree of two-dimensionality of the flow \red{and counter} the formation of LSVs. Nevertheless, numerical studies using parametrized Ekman pumping \red{BCs report} an upscale transfer of kinetic energy that \red{results in intermittent LSV formation} with \red{irregular} intensity and coherence at $Pr=1$ \cite{julien2016nonlinear,plumley2016effects}. The Prandtl number $Pr=\nu / \kappa$ \red{describes the diffusive properties of the fluid, where $\nu$ and $\kappa$ respectively are its kinematic viscosity and thermal diffusivity}.

%  “In this letter”: Summary of results and broad impact.

In this \red{paper}, we demonstrate the presence of coherent, long-lived \red{LSVs} in rotationally-constrained RRBC despite Ekman pumping interference from the boundary layers \red{(BLs)}. And we do so \red{for fluids of} different Prandtl \red{number}: low $Pr=0.1$, \red{relevant to liquid metals as in the Earth's outer core}, and high $Pr=5$, \red{relevant to oceanic processes and laboratory experiments with water}. We \red{identify the direct energy transfer from the smaller scales to the largest scale in the domain,} while rotation subdues any significant disturbances from the \red{BLs}. \red{Our observation of LSVs in domains with no-slip plates opens up laboratory modeling of LSV formation, a process that is omnipresent in large-scale natural flows.}

RRBC \red{is governed} by three nondimensional parameters: the Rayleigh number $Ra = g\alpha\Delta\theta H^3 / \nu \kappa$ that quantifies the intensity of the buoyant forcing, the Ekman number $Ek = \nu / 2 \Omega H^2$ that measures the (inverse) strength of rotation\red{,} and the Prandtl number $Pr$. Here, \red{$g$ is the gravitational acceleration, $\alpha$ the thermal expansion coefficient of the fluid,} $\Delta \theta$ and $H$ are the temperature difference and distance between the top and bottom \red{plates}, respectively, \red{and} $\Omega$ is the rotation rate\red{.}

We simulate \red{the Navier--Stokes and heat equations for an incompressible fluid} with NS boundaries at $Pr=0.1$, $Ra=1\times10^{10}$ and $2\times10^{-7}\leq Ek\leq6\times10^{-6}$; at $Pr=5.5$, $Ek=3\times 10^{-7}$ and $5.5\times10^{9}\leq Ra\leq2\times10^{10}$; and at $Pr=5.2$, $Ek=1\times 10^{-7}$ and $3\times10^{10}\leq Ra\leq1.5\times10^{12}$ \footnote{The slight difference in $Pr$ between the two simulation series is for comparison with (ongoing) experiments in our group \cite{cheng2018heuristic}.}. We also simulate the flow at $Pr=0.1$, $Ra=1\times10^{10}$ and $Ek=2.5\times10^{-7}$ with SF \red{BCs} for comparison with the NS case. We \red{employ a} second-order \red{accurate} finite-difference \red{discretization} \cite{ostilla2015multiple} \red{on} a grid vertically denser near the \red{plates} to appropriately resolve the thinner (Ekman or thermal) BL with a minimum of $10$ points. The \red{dimensions} of the \red{horizontally periodic} computational domain \red{are} $10L_c\times10L_c\times1$, where $L_c$ is the horizontal wavelength \red{most unstable for} onset of convection \cite{chandrasekhar2013hydrodynamic}. At $Pr=0.1$, we set resolutions up to $1408\times1408\times1280$ points to resolve the Kolmogorov length scale $\eta_K$\red{, the smallest active length scale in the flow}. At $Pr=5$, the Batchelor scale $\eta_B\red{=\eta_K/Pr^{1/2}}$ \red{(smallest active length scale in the temperature field)} is smaller than $\eta_K$ and we use a multiple-resolution approach\red{:} a fine grid \red{for the temperature field} resolves \red{down to} $\eta_B$ and a coarser grid \red{is used for the velocity field. This} alleviates the computational requirements and allow\red{s} us to explore different values of $Ra$ at a very low $Ek$. \red{W}e use up to $1536\times1536\times2048$ points to resolve $\eta_B$. \red{We present our results as a function of the supercriticality $R\equiv Ra/Ra_c$, where $Ra_c$ is the critical Rayleigh number for the onset of oscillatory (stationary) convection at $Pr<0.68$ ($Pr\ge 0.68$) \footnote{The critical Rayleigh number for onset of oscillatory convection ($Pr<0.68$) is $Ra_c=17.4(Ek/Pr)^{-4/3}$; for onset of steady convection ($Pr\ge 0.68$) it is $Ra_c=8.7Ek^{-4/3}$ \cite{chandrasekhar2013hydrodynamic}.}.}

\begin{figure}[b]
    \includegraphics[trim={.15in .22in .15in .15in},clip,width=.48\textwidth]{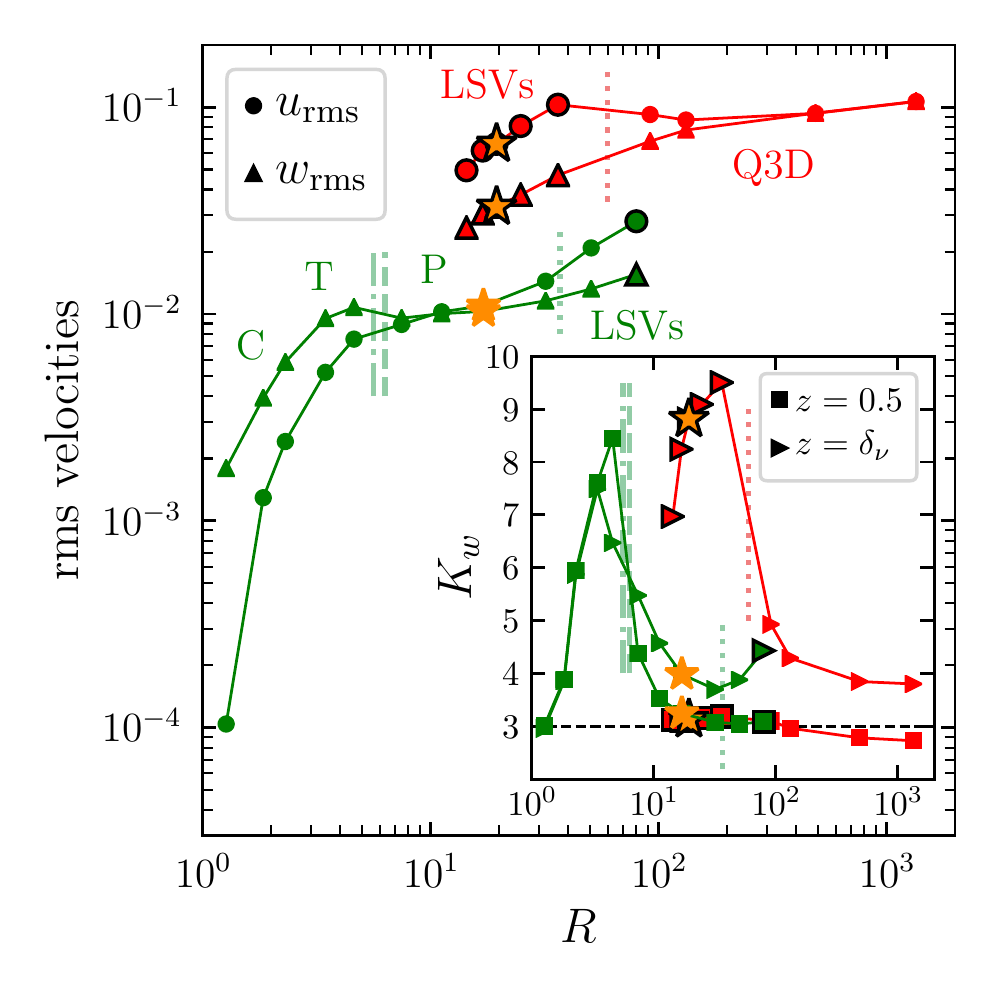} % left lower right upper
    \caption{Root-mean-square horizontal (circles) and vertical (up triangles) velocities at mid-height \red{($z=0.5$)} at $Pr=0.1$ (red symbols) and $5$ (green symbols) as a function of $R\red{=} Ra/Ra_c$. Inset: kurtosis \red{of vertical velocity $K_w$} at mid-height (\red{$z=0.5$;} squares) and at the Ekman BL \red{thickness} (\red{$z=\delta_\nu$;} right triangles). The \red{horizontal} dashed line at $K_w=3$ indicates Gaussian kurtosis. \red{Vertical d}ash-dotted and dashed green lines are the predicted transitions from convective columns (T) to plumes (P) in Refs. \cite{cheng2015laboratory} and \cite{nieves2014statistical}, respectively, and the \red{vertical} dotted red lines are our estimated transitions to LSVs. Symbols with black edges \red{represent LSV flow states}. The orange stars are the cases \red{selected for further analysis and comparison.}}
    \label{fig:velrms_wkurt}
\end{figure}

\cref{fig:velrms_wkurt} \red{presents the} root-mean-square \red{(rms) values of horizontal ($u_\mathrm{rms}$) and vertical ($w_\mathrm{rms}$) velocity as well as} the kurtosis of the vertical velocity $K_w$ as a function of \red{$R$.} In the bulk, the relative magnitude of horizontal to vertical \red{velocity} fluctuations exhibits significantly different behavior specific of the flow \red{structure (as identified by \cite{julien2012statistical})}: for steady cells (C) and convective Taylor columns (T) vertical fluctuations are stronger than horizontal \red{fluctuations}, for plumes (P) (and in Q3D turbulence at $Pr=0.1$) the\red{y} are comparable\red{, while} for LSVs the horizontal fluctuations \red{are larger} as a sign of the two-dimensionali\red{zation} of the flow.

Values of kurtosis larger than that of a Gaussian distribution (i.e. $K_w>3$) are associated with increased likelihood of strong vertical velocity \red{\cite{julien2012statistical} and are indicative of the presence of coherent structures}. \red{Ref. \cite{julien2012statistical} report $K_w>3$ for cells, columns, and plumes, while in the so-called geostrophic turbulence state (where LSVs are observed) $K_w$ is reduced to $3$ again as in homogeneous isotropic turbulence. Our observations for the bulk are the same (at mid-height $z=0.5$).} Both the larger horizontal-to-vertical velocities and the Gaussian kurtosis in the bulk are clear signatures of the Q2D turbulent state of the flow. \red{The kurtosis $K_w$ computed close to the plate at height $z=\delta_\nu$, the Ekman BL thickness based on the peak position of $u_\mathrm{rms}$, follows the bulk trend for most of the $R$ range, except for in the LSV state where it retains values larger than $3$. Thus there are still smaller-scale coherent structures formed by the Ekman BLs, though they cannot break the LSVs.} This observation will be further discussed later on together with the kinetic energy budget near the walls. \red{We shall first consider} the bulk phenomenology.

\begin{figure}
    \subfloat[$Pr=0.1$, $R=20$ and NS]{
        \includegraphics[width=.23\textwidth]{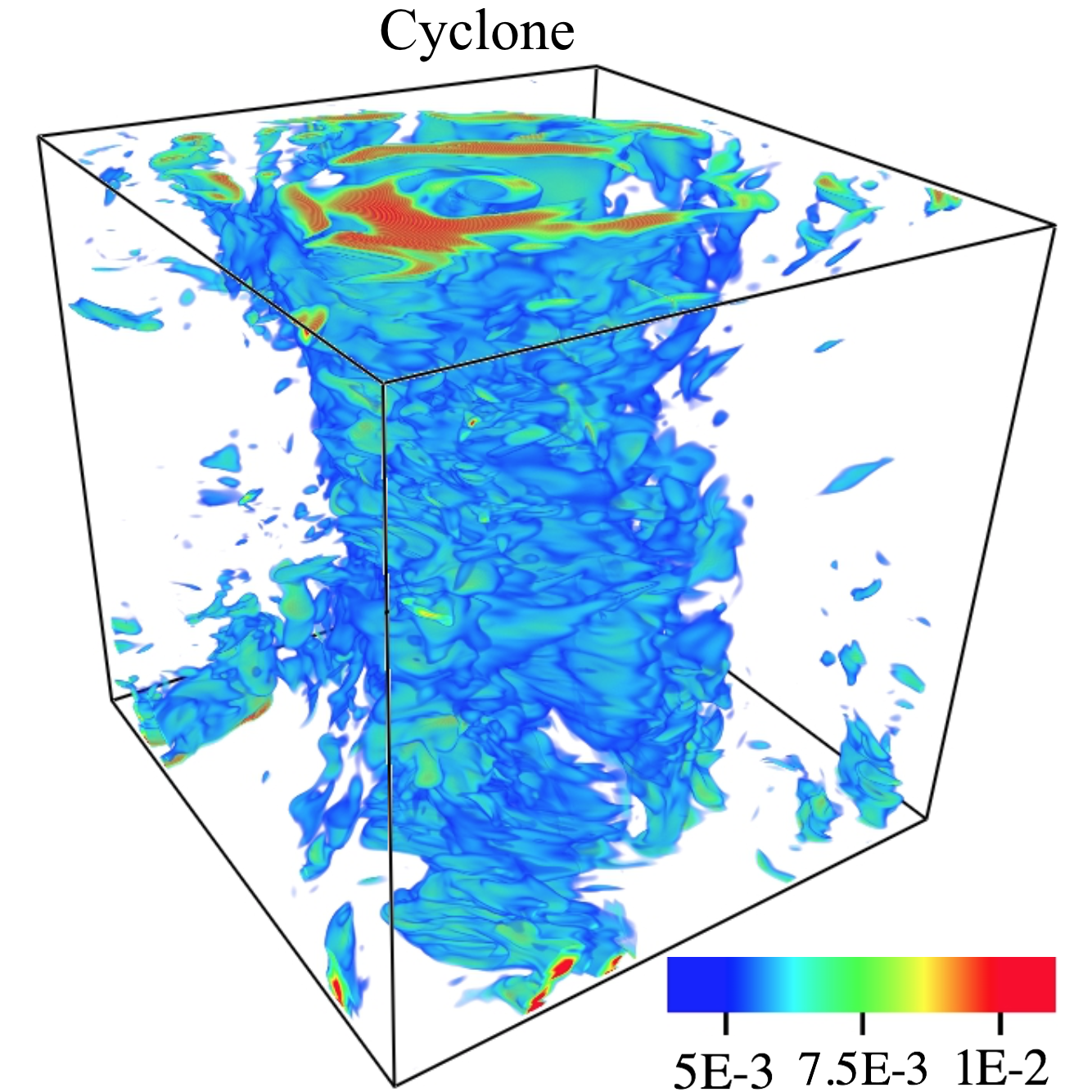}
        \label{fig:rendering_pr0p1_lsv}}
    \subfloat[$Pr=5.2$, $R=80$ and NS]{
        \includegraphics[width=.23\textwidth]{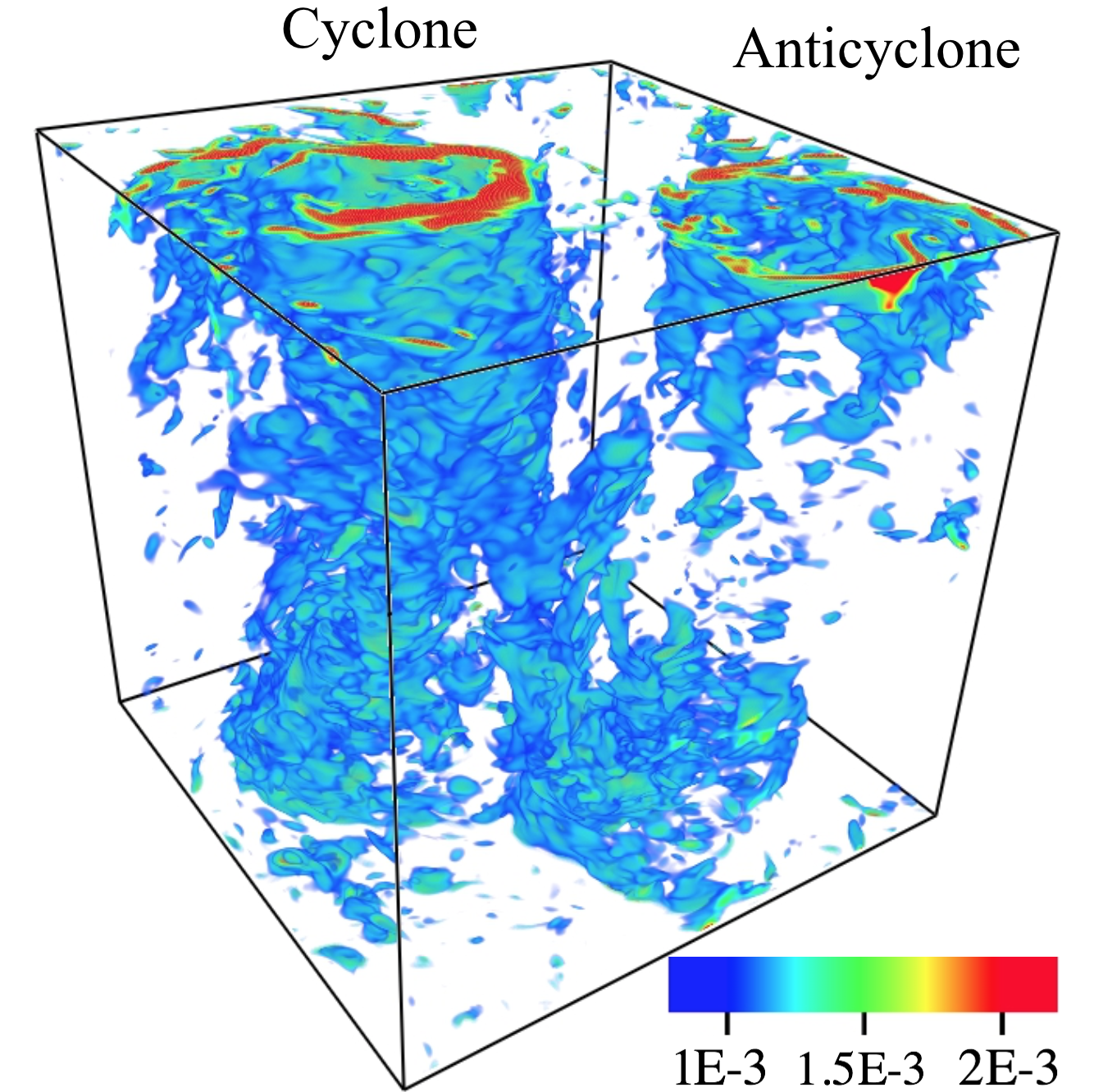}
        \label{fig:rendering_pr5p2_lsv}}
        
    \subfloat[$Pr=5.2$, $R=11$ and NS]{
        \includegraphics[width=.23\textwidth]{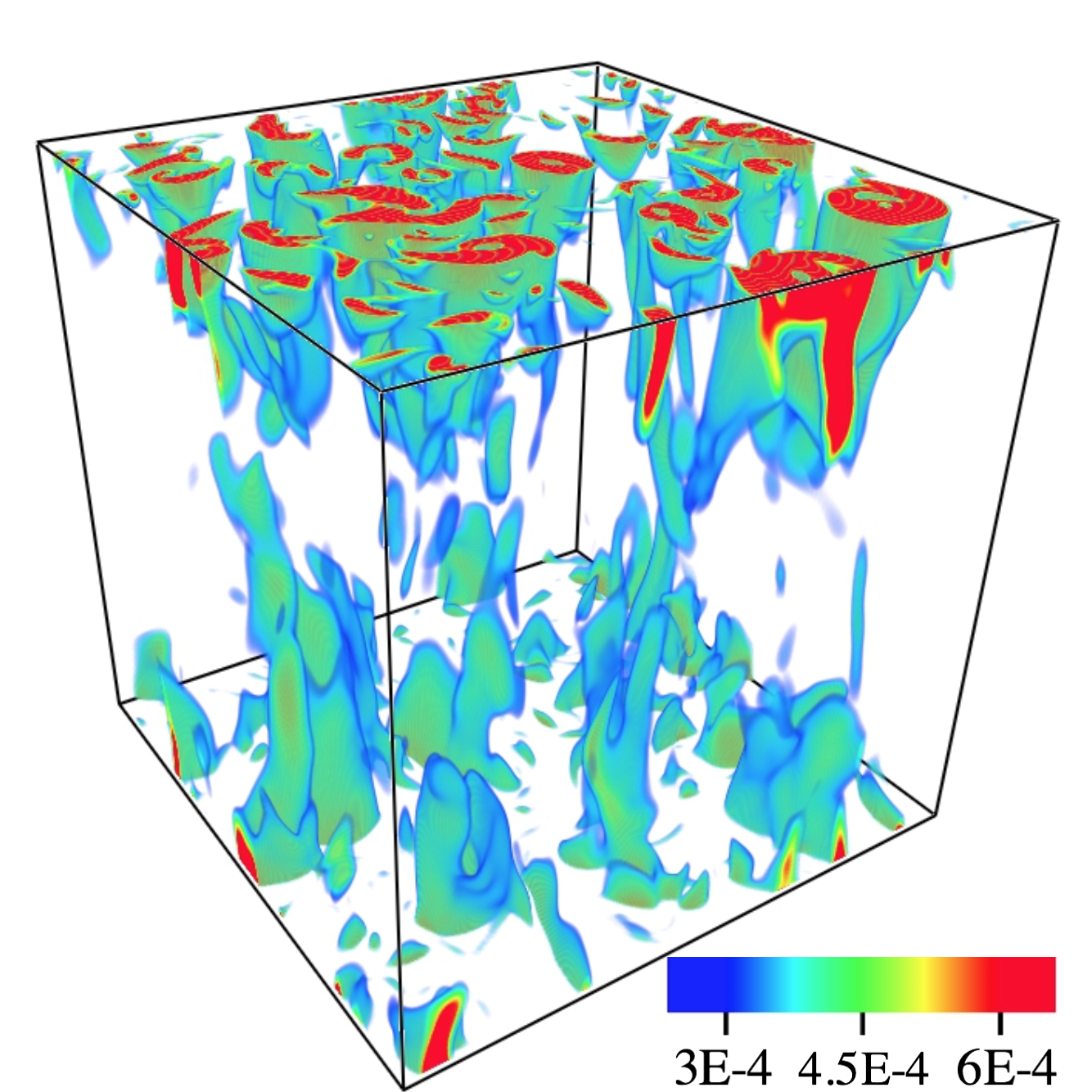}
        \label{fig:rendering_pr5p2_plumes}}
    \subfloat[$Pr=0.1$, $R=20$ and SF]{
        \includegraphics[width=.23\textwidth]{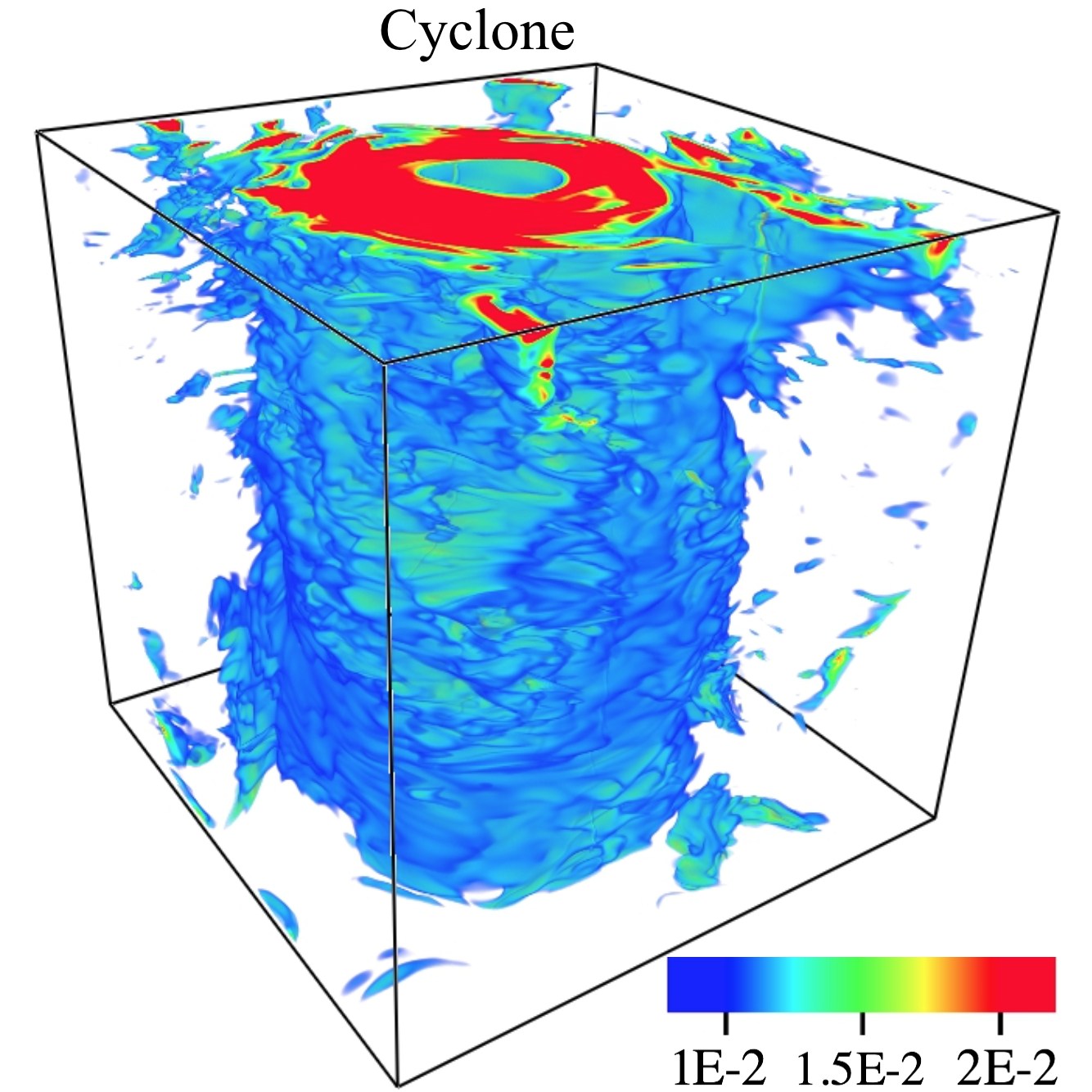}
        \label{fig:rendering_pr0p1_sf}}
         
    \caption{Snapshots of the horizontal kinetic energy \red{for four different cases in terms of Prandtl number, supercriticality and boundary condition.} Cases (a-c) are indicated in \cref{fig:velrms_wkurt} with orange stars. \red{Case} (d) is at the same $Pr$, $Ra$ and $Ek$ (and same $R$) as (a), but with SF \red{BCs}. For clarity, the domains are stretched horizontally by a factor of (a,d) $3.1$ and (b,c) $4.5$.}
    \label{fig:rendering}
\end{figure}

\red{T}he flow at $Pr=0.1$ and $R=20$\red{, visualized in Fig. \ref{fig:rendering}(a),} consists of one large-scale coherent structure with cyclonic \red{(positive)} vertical vorticity that extends over the entire \red{height} of the domain. Our comparative case at the same $Pr$ and $R$ \red{but} with SF \red{BCs (\cref{fig:rendering_pr0p1_sf})} exhibits a similar flow morphology. The vortex is embedded in a\red{n environment with} weakly anticyclonic \red{(negative) vorticity}. Remarkably, at $Pr=5.2$ and $R=80$ \red{a dipole consisting of} both \red{a} cyclonic and \red{an} anticyclonic \red{vortex is} present (left and right vortices in \red{Fig.\ref{fig:rendering}(b)}, respectively). \red{Just a}s the vortex monopole, the dipole spans the \red{height of the} domain. For the duration of our simulations, the vortices \red{have shown} to be long-lived structures without significant horizontal displacement. At $Pr=5.2$ and lower supercriticality $R=11$, the upscale energy transfer cannot develop, the flow does not organize into large coherent structures and instead exhibits plumes of smaller scale (see \cref{fig:rendering_pr5p2_plumes}) \cite{kunnen2010experimental,kunnen2010vortex,julien2012statistical}.
 
\red{Figs. \ref{fig:rendering}(a,b,d)} show that the flow favors cyclonic vortices over anticyclones. \red{Whenever an anticyclone is present (as in Fig. \ref{fig:rendering}b), it tends to be weaker than its cyclonic counterpart.} The cyclon\red{e--}anticyclon\red{e} asymmetry in rotating flows has been extensively discussed in many numerical and experimental studies\red{, e.g., Refs.} \cite{hopfinger1982turbulence, baroud2003scaling, sreenivasan2008formation, godeferd2015structure}. In the case of vortex condensation, the \red{presence or absence of the anticyclone is postulated to be the result of either of two} saturation mechanism\red{s} \red{\cite{seshasayanan2018condensates}}: \red{in a so-called ``viscous condensate" (as our \cref{fig:rendering_pr5p2_lsv}) the viscous dissipation at large scales is large enough to match the upscale energy transfer, while in a so-called ``rotating condensate" (as our \cref{fig:rendering_pr0p1_lsv}) the amplitude of the anticyclone becomes so large that it locally cancels the conditions for Q2D flow, leading to a 3D turbulent downscale transfer.}

Spectral condensation of kinetic energy at large scales is responsible for the formation of large-scale structures in turbulence. To study the (rate of) transfer of energy among scales, we use the shell-to-shell energy transfer function \cite{alexakis2005shell,mininni2009scale,favier2014inverse}
\begin{equation}
T(Q,K) \equiv - \int_{\red{V}} \bm{u}_K ( \bm{u} \bm{\cdot}  \bm{\nabla} )  \bm{u}_Q \; d \bm{x}^3
\label{eq:energy_transfer}
\end{equation}
which \red{describes the energy transfer from the Fourier-filtered flow field $\bm{u}_Q$ of wavenumber $Q$ to the filtered field of wavenumber $K$ (Fourier transforms performed in the horizontal periodic directions; ring-like shells of different horizontal wavenumbers are selected; integration over the simulation volume $V$)}. If $T(Q,K)>0$, \red{the mode with wavenumber} $Q$ transfer\red{s} energy to \red{mode} $K$ via triadic interactions, \red{while mode} $Q$ receives energy from \red{mode} $K$ when $T(Q,K)<0$. The energy transfer function is antisymmetric and \red{can be derived} from the global budget equation of the modal kinetic energy \red{\cite{alexakis2005shell,mininni2009scale,favier2014inverse}}. The integration along the vertical direction does not discern \red{bulk and BL} region\red{s}. \red{Since in all considered cases} the BLs account for less than 2\% of the volume\red{,} \cref{eq:energy_transfer} provides a \red{good} description of \red{spectral energy transfer in the bulk}.

\begin{figure}
    \subfloat[$Pr=0.1$, $R=20$ and NS]{
        \includegraphics[width=.23\textwidth,trim=.8in 2.7in 1.1in 2.6in,clip]{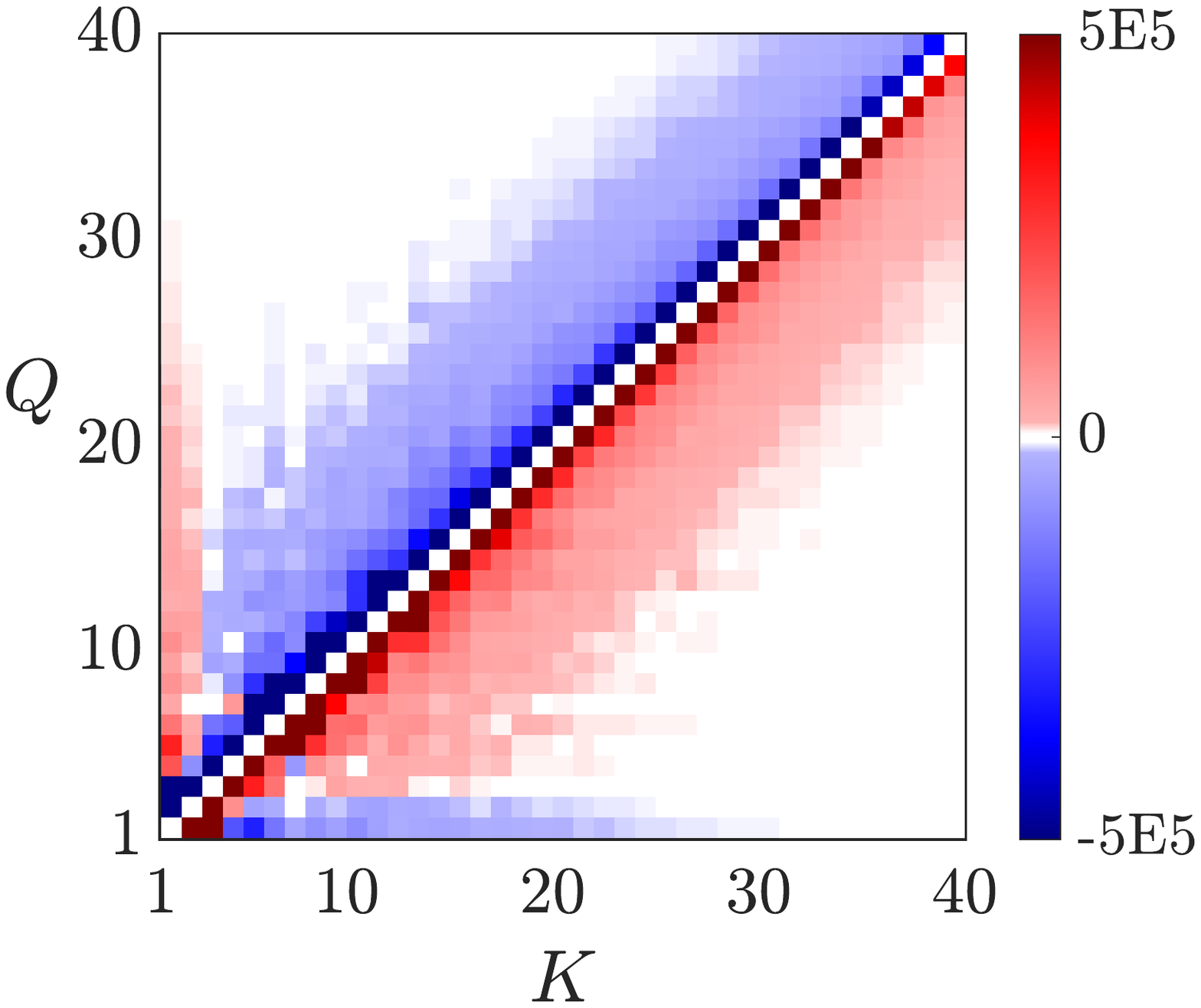} % left bottom right top
        \label{fig:transfer_pr0p1_lsv}}
    \subfloat[$Pr=5.2$, $R=80$ and NS]{
        \includegraphics[width=.23\textwidth,trim=.8in 2.7in 1.1in 2.6in,clip]{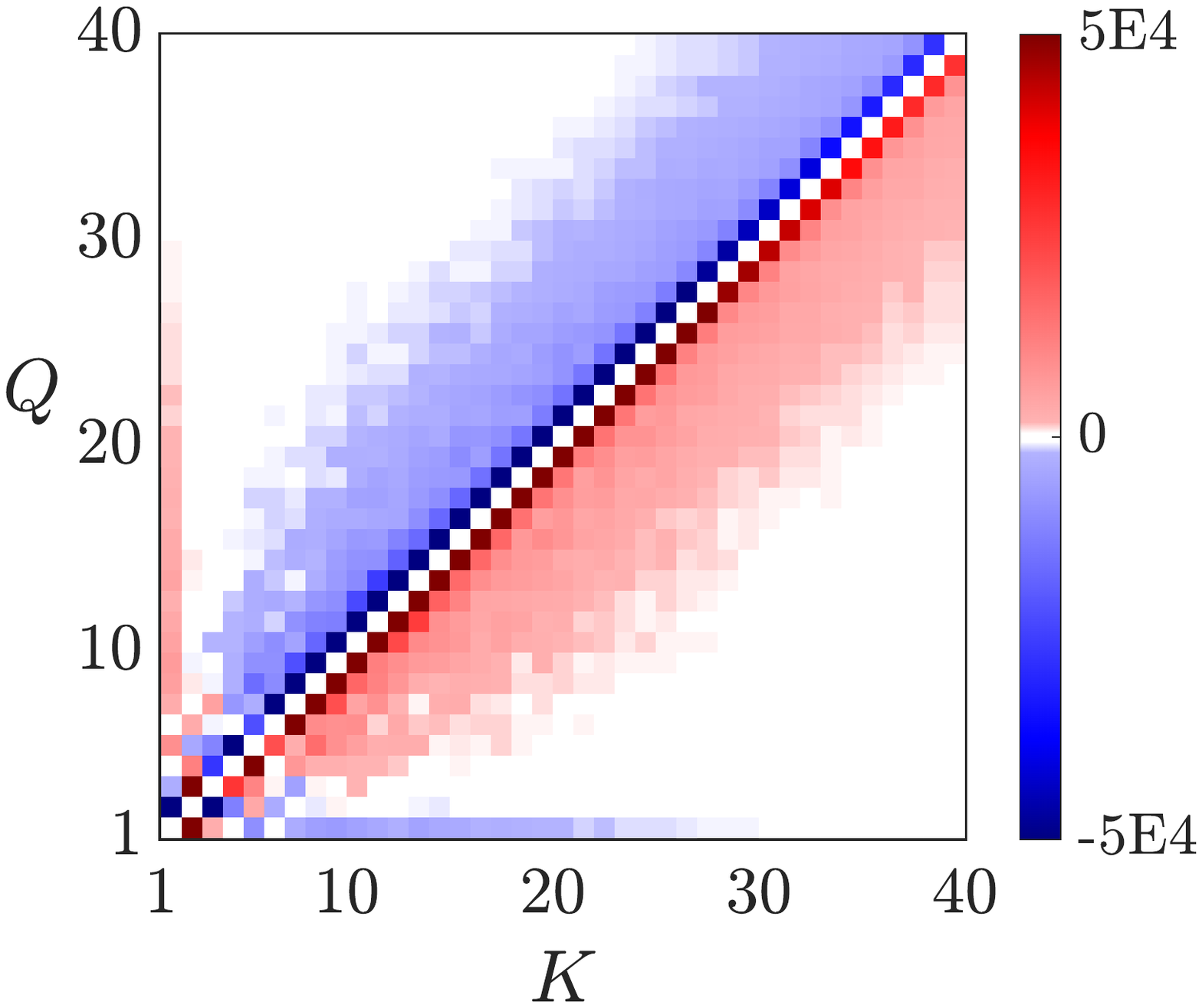}
         \label{fig:transfer_pr5p2_lsv}}
         
    \subfloat[$Pr=5.2$, $R=11$ and NS]{
        \includegraphics[width=.23\textwidth,trim=.8in 2.7in 1.1in 2.6in,clip]{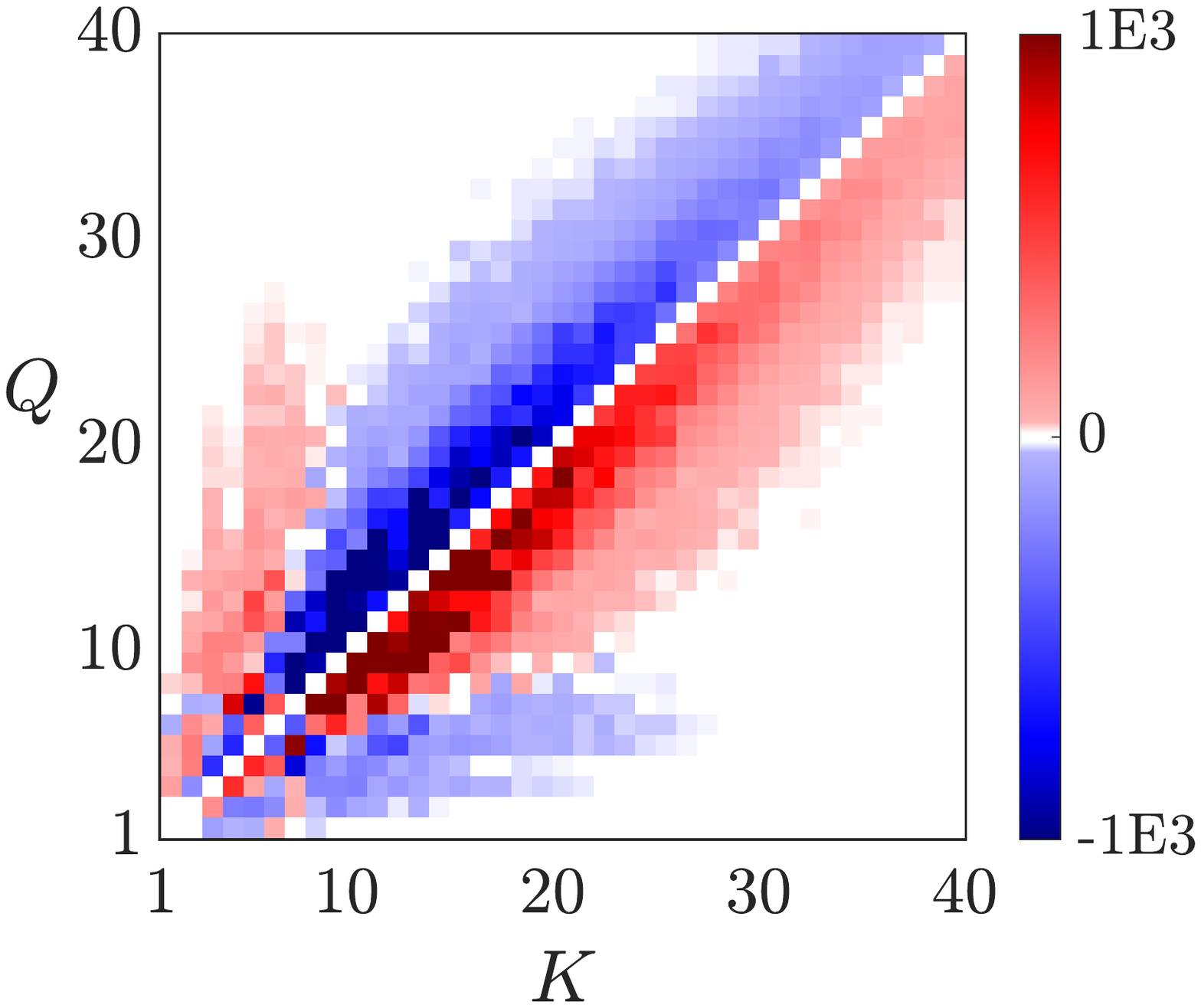}
         \label{fig:transfer_pr5p2_plumes}}
    \subfloat[$Pr=0.1$, $R=20$ and SF]{
        \includegraphics[width=.23\textwidth,trim=.8in 2.7in 1.1in 2.6in,clip]{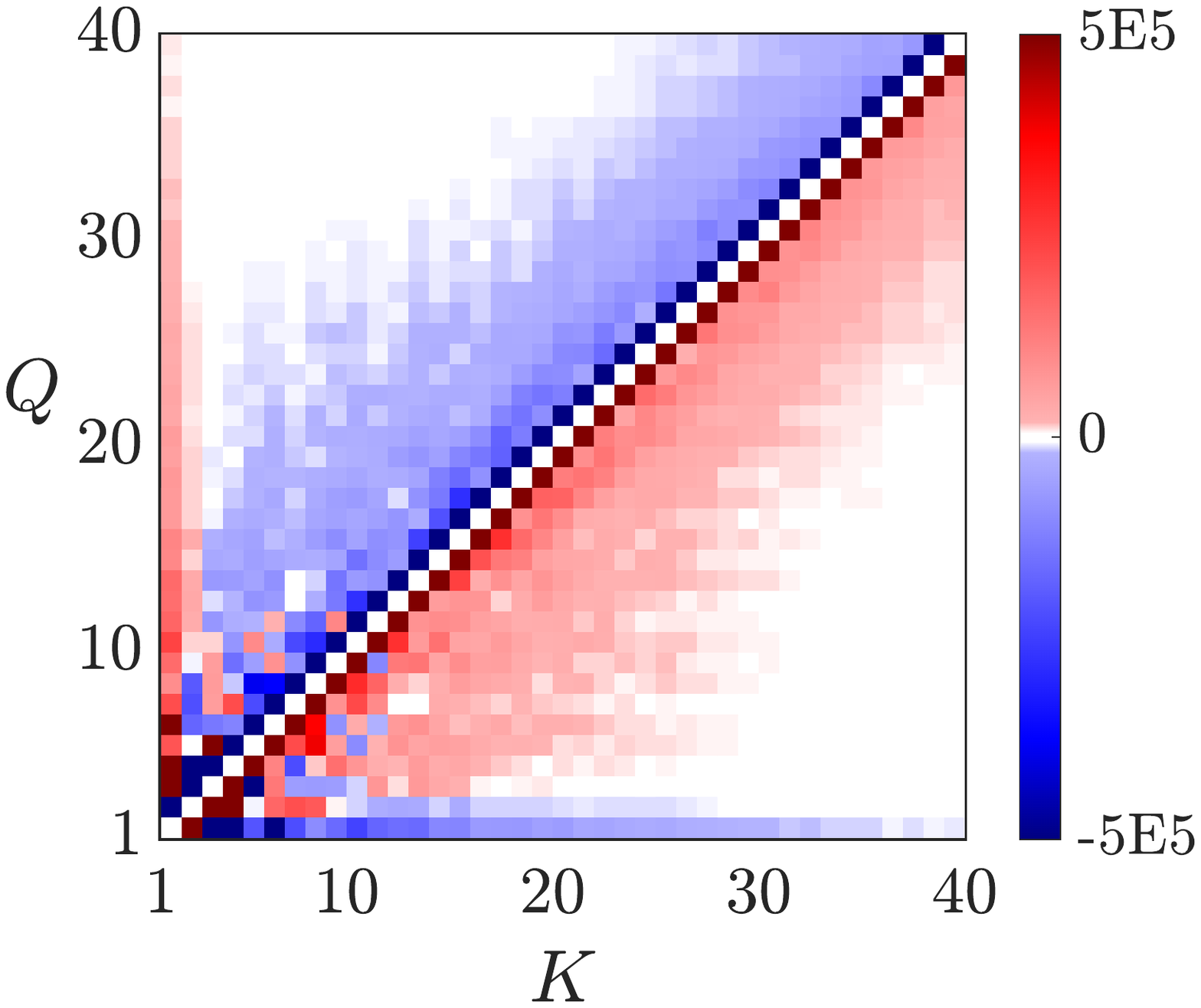}
         \label{fig:transfer_pr0p1_sf}}
         
    \caption{\red{S}hell-to-shell energy transfer function $T(Q,K)$ \red{for the four selected cases.} \red{Note} the \red{different} limits of the color bar for each instance\red{; t}he color scale is chosen to highlight the main energy transfers.}
    \label{fig:transfer}
\end{figure}

In all cases displayed in Fig. \ref{fig:transfer} the $Q=K-1$ diagonal in $T(Q,K)$ is positive, representative of a spectrally local downscale transfer of kinetic energy\red{, i.e., a given mode $Q$ predominantly transfers energy to mode $K=Q+1$}. On the other hand, \red{Figs. \ref{fig:transfer}(a,b,d)} reveal an \emph{upscale} transfer in the presence of LSVs: at $Q=1$ ($K=1$) and over a wide range of values of $K$ ($Q$), $T(Q,K)$ is mostly negative (positive). This indicates a \red{spectrally nonlocal} upscale transfer, i.e. the small scales directly transfer energy to the large\red{st scale} without the participation of intermediate scales. Both downscale and upscale transfers of kinetic energy coexist in the flow. As expected, a downscale energy transfer dominates in the plumes \red{flow state and no direct transfer to the lowest wavenumber mode is found (\cref{fig:transfer_pr5p2_plumes})}.

The \red{dual} energy transfer is therefore a leading feature of the LSVs. However, \red{we know that the boundary layers can decisively affect the structuring of the bulk flow.} To \red{assess the role of the BLs} we \red{calculate} the height-dependent planar budget of kinetic energy $E=\red{\tfrac{1}{2}(u^2+v^2+w^2)}$ \cite{deardorff1967investigation,kerr2001energy,kunnen2009turbulence}\red{:}
\begin{equation}
 \left\langle w\theta  \right\rangle
- \frac{\partial}{\partial z}  \left\langle w E  \right\rangle
- \frac{\partial}{\partial z}  \left\langle w p  \right\rangle \\
+ 2 \sqrt{\frac{Pr}{Ra}} \frac{\partial}{\partial z}  \left\langle u_i s_{i3}  \right\rangle
-  \left\langle \epsilon  \right\rangle = 0
\label{eq:budget}
\end{equation}
where $\theta$ is temperature, $p$ is pressure, $s_{ij} = \red{\tfrac{1}{2}} \left( \partial_j u_i +  \partial_i u_j \right)$ is the deformation rate tensor \red{(summation over repeated indices implied;} $i=1,2$ are the \red{horizontal directions} and \red{$i=3$ the vertical direction}) and $\langle \cdot  \rangle$ indicates averaging over the horizontal directions and in time. The first term in \cref{eq:budget} is buoyant production, the second, third and fourth terms are turbulent \red{transport}, pressure \red{transport} and viscous transport, respectively, and the fifth term is dissipation of kinetic energy $\langle \epsilon \rangle = 2 \sqrt{Pr/Ra} \left\langle  s_{ij} s_{ij}  \right\rangle$ \cite{pope2001turbulent}.

\begin{figure*}
    \includegraphics[width=\textwidth]{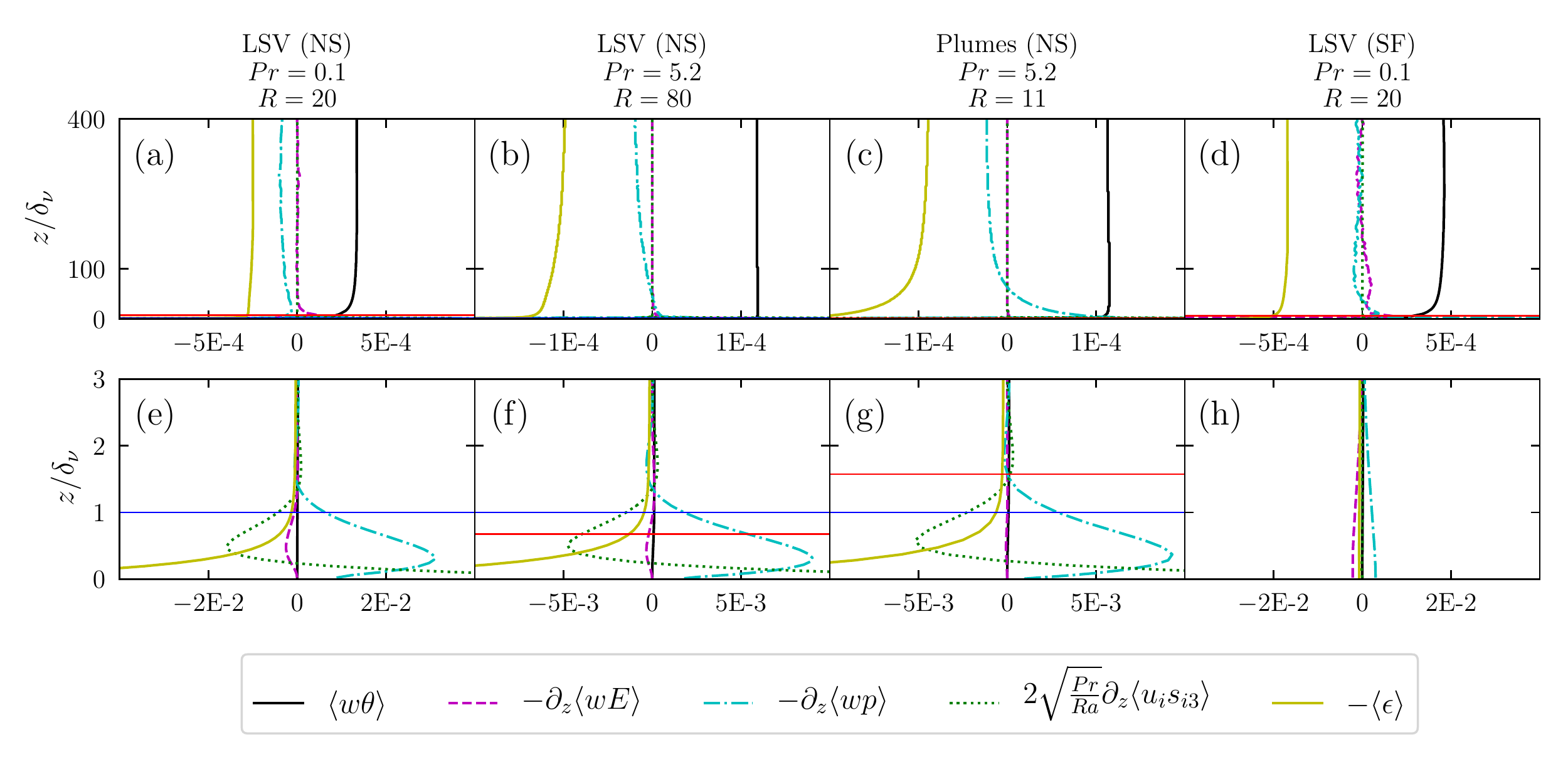} % left bottom right top ,trim=.27in .28in .22in .2in,clip
    \caption{\red{Vertical profiles of k}inetic energy budget \red{terms (Eq. \ref{eq:budget})} (a-d) \red{in} the bulk and (e-h) close to the bottom wall. \red{Note} the change in scale of the horizontal axes for the different panels. The vertical coordinate is rescaled by the \red{corresponding} viscous BL thickness $\delta_\nu=\mathcal{O}(Ek^{1/2})$, except for the SF case where this BL is absent and we use the $\delta_\nu$ of its NS counterpart. All profiles are symmetric about mid-height ($z/\delta_\nu\sim400$). The blue and red \red{horizontal} lines depict the viscous and thermal BLs, respectively. The latter is outside the plotting interval in (e) at $z/\delta_\nu=8$ and in (h) at $z/\delta_\nu=6$.}
    \label{fig:budget_bulk_bl}
\end{figure*}

In the bulk of the LSVs and the plumes with no-slip condition \red{(Figs. \ref{fig:budget_bulk_bl}a-c)} buoyan\red{t} production provides the kinetic energy \red{that is in large part dissipated;} pressure alone transports the rest towards the BL. The other two transport mechanisms are marginal in \red{the bulk}. Interestingly, \red{for the LSV case with SF BCs,} production is \red{practically} balanced by dissipation, and the pressure transport is mostly absent throughout the bulk. Thus, the transport due to pressure fluctuations is \red{driven by the boundary layers (Ekman pumping) but} does not \red{prevent} the \red{upscale} transfer of kinetic energy.

Close to the walls \red{(Figs. \ref{fig:budget_bulk_bl}e-g)} the leading terms of the budget are the pressure transport from the bulk, the viscous transport from the BL edge closer to the wall and the dissipation in this region. These terms are at least one order of magnitude larger than in the bulk, but \red{largely} confined to the viscous BL. Therefore, as rotation hinders the plumes throughout the bulk, the magnitude of the budget terms is also heavily restrained. Viscosity, on the other hand, strengthen them in the BL through Ekman pumping and redistributes the budget terms.  A comparison of the energy budget for the NS and SF cases at $Pr=0.1$ \red{(Figs. \ref{fig:budget_bulk_bl}e and h}, respectively) further accentuates the role of the frictional viscous BL in the enhancement of the flow dynamics close to the walls.

% Conclusions

To conclude, we \red{demonstrate} the presence of large-scale vortices in horizontally periodic RRBC with no-slip \red{boundary conditions at} top and bottom \red{for both} $Pr=0.1$ and $5$. We \red{show} that the formation of such long-lived coherent structures is due to spectral condensation of the kinetic energy transferred upscale; a process that is due to the two-dimensionalization of the flow by strong rotation.

The energy accumulates at the largest horizontal scale available, i.e. the domain width (at $Q=1$ in \cref{fig:transfer_pr0p1_lsv,fig:transfer_pr5p2_lsv}), indicating that spectral condensation is \red{bounded by the} confinement, \red{not by friction induced by the Ekman boundary layers. In our domains with small width-to-height ratios condensation is halted by confinement rather than by} wall dissipation of the LSV energy at finite rotation rates. Instead, the \red{boundary layers postpone the occurrence of upscale energy transfers to significantly higher rotation rates (at the same supercriticality) than for stress-free boundary conditions.} \red{The boundary layers also facilitate} the formation of plumes through Ekman pumping, although we \red{show} that these vertical flows do not penetrate into the Q2D bulk\red{.} Moreover, the establishment of strong horizontal \red{motions} in the bulk \red{induces a mean shear \cite{xia2011upscale} that} protects the bulk from the boundary\red{-}layer plumes and \red{as such} contributes to the two-dimensionalization of the \red{flow.} The role of the \red{Prandtl number (}fluid properties\red{)} on the susceptibility of the flow to develop LSVs, and to favor \red{either} saturation mechanism \red{\cite{seshasayanan2018condensates}}, is subject \red{to} further investigation. \red{In SF simulations it appears that geostrophic turbulence (and LSV formation) is more easily reached at smaller $Pr$ \cite{julien2012statistical}; this finding is replicated here for NS simulations. We postulate that coherent plumes, formed by the Ekman BLs, with strong temperature contrast, are longer-lived at high $Pr$ given weaker thermal diffusion. Hence they can more proficiently disturb LSV formation.}

Large-scale flow \red{organization is an} ubiquitous \red{feature} in geophysic\red{al} and astrophysic\red{al flows}. \red{Our study identifies that upscale transfer is delayed by Ekman boundary layers but it can still occur at more extreme parameter values. This opens up laboratory modeling of these flows in the} rotating Rayleigh\red{--}B\'enard configuration\red{ \cite{cheng2018heuristic}. We show} that no-slip boundaries \red{promote} the development of small-scale structures in the flow, \red{but given strong enough rotational constraint the upscale energy transfer prevails.}

\begin{acknowledgments}
\red{A.J.A.G., M.M., J.S.C., and R.P.J.K.} received funding from the European Research Council (ERC) under the European Union’s Horizon 2020 research and innovation programme (Grant agreement No. 678634). We are grateful for the support of the Netherlands Organisation for Scientific Research (NWO) for the use of supercomputer facilities (Cartesius) under Grants No. 15462, 16467 and 2019.005. Volume renders are produced using VAPOR (www.vapor.ucar.edu), a product of the Computational Information Systems Laboratory at the National Center for Atmospheric Research.
\end{acknowledgments}


\begin{thebibliography}{52}%
\makeatletter
\providecommand \@ifxundefined [1]{%
 \@ifx{#1\undefined}
}%
\providecommand \@ifnum [1]{%
 \ifnum #1\expandafter \@firstoftwo
 \else \expandafter \@secondoftwo
 \fi
}%
\providecommand \@ifx [1]{%
 \ifx #1\expandafter \@firstoftwo
 \else \expandafter \@secondoftwo
 \fi
}%
\providecommand \natexlab [1]{#1}%
\providecommand \enquote  [1]{``#1''}%
\providecommand \bibnamefont  [1]{#1}%
\providecommand \bibfnamefont [1]{#1}%
\providecommand \citenamefont [1]{#1}%
\providecommand \href@noop [0]{\@secondoftwo}%
\providecommand \href [0]{\begingroup \@sanitize@url \@href}%
\providecommand \@href[1]{\@@startlink{#1}\@@href}%
\providecommand \@@href[1]{\endgroup#1\@@endlink}%
\providecommand \@sanitize@url [0]{\catcode `\\12\catcode `\$12\catcode
  `\&12\catcode `\#12\catcode `\^12\catcode `\_12\catcode `\%12\relax}%
\providecommand \@@startlink[1]{}%
\providecommand \@@endlink[0]{}%
\providecommand \url  [0]{\begingroup\@sanitize@url \@url }%
\providecommand \@url [1]{\endgroup\@href {#1}{\urlprefix }}%
\providecommand \urlprefix  [0]{URL }%
\providecommand \Eprint [0]{\href }%
\providecommand \doibase [0]{https://doi.org/}%
\providecommand \selectlanguage [0]{\@gobble}%
\providecommand \bibinfo  [0]{\@secondoftwo}%
\providecommand \bibfield  [0]{\@secondoftwo}%
\providecommand \translation [1]{[#1]}%
\providecommand \BibitemOpen [0]{}%
\providecommand \bibitemStop [0]{}%
\providecommand \bibitemNoStop [0]{.\EOS\space}%
\providecommand \EOS [0]{\spacefactor3000\relax}%
\providecommand \BibitemShut  [1]{\csname bibitem#1\endcsname}%
\let\auto@bib@innerbib\@empty
%</preamble>
\bibitem [{\citenamefont {Charney}(1971)}]{charney1971geostrophic}%
  \BibitemOpen
  \bibfield  {author} {\bibinfo {author} {\bibfnamefont {J.~G.}\ \bibnamefont
  {Charney}},\ }\href@noop {} {\bibfield  {journal} {\bibinfo  {journal} {J.
  Atmos. Sci.}\ }\textbf {\bibinfo {volume} {28}},\ \bibinfo {pages} {1087}
  (\bibinfo {year} {1971})}\BibitemShut {NoStop}%
\bibitem [{\citenamefont {Pedlosky}(1979)}]{pedlosky2013geophysical}%
  \BibitemOpen
  \bibfield  {author} {\bibinfo {author} {\bibfnamefont {J.}~\bibnamefont
  {Pedlosky}},\ }\href@noop {} {\emph {\bibinfo {title} {Geophysical Fluid
  Dynamics}}}\ (\bibinfo  {publisher} {Springer},\ \bibinfo {year}
  {1979})\BibitemShut {NoStop}%
\bibitem [{\citenamefont {Smith}\ and\ \citenamefont
  {Waleffe}(1999)}]{smith1999transfer}%
  \BibitemOpen
  \bibfield  {author} {\bibinfo {author} {\bibfnamefont {L.~M.}\ \bibnamefont
  {Smith}}\ and\ \bibinfo {author} {\bibfnamefont {F.}~\bibnamefont
  {Waleffe}},\ }\href@noop {} {\bibfield  {journal} {\bibinfo  {journal} {Phys.
  Fluids}\ }\textbf {\bibinfo {volume} {11}},\ \bibinfo {pages} {1608}
  (\bibinfo {year} {1999})}\BibitemShut {NoStop}%
\bibitem [{\citenamefont {Campagne}\ \emph {et~al.}(2014)\citenamefont
  {Campagne}, \citenamefont {Gallet}, \citenamefont {Moisy},\ and\
  \citenamefont {Cortet}}]{campagne2014direct}%
  \BibitemOpen
  \bibfield  {author} {\bibinfo {author} {\bibfnamefont {A.}~\bibnamefont
  {Campagne}}, \bibinfo {author} {\bibfnamefont {B.}~\bibnamefont {Gallet}},
  \bibinfo {author} {\bibfnamefont {F.}~\bibnamefont {Moisy}},\ and\ \bibinfo
  {author} {\bibfnamefont {P.-P.}\ \bibnamefont {Cortet}},\ }\href@noop {}
  {\bibfield  {journal} {\bibinfo  {journal} {Phys. Fluids}\ }\textbf {\bibinfo
  {volume} {26}},\ \bibinfo {pages} {125112} (\bibinfo {year}
  {2014})}\BibitemShut {NoStop}%
\bibitem [{\citenamefont {Godeferd}\ and\ \citenamefont
  {Moisy}(2015)}]{godeferd2015structure}%
  \BibitemOpen
  \bibfield  {author} {\bibinfo {author} {\bibfnamefont {F.~S.}\ \bibnamefont
  {Godeferd}}\ and\ \bibinfo {author} {\bibfnamefont {F.}~\bibnamefont
  {Moisy}},\ }\href@noop {} {\bibfield  {journal} {\bibinfo  {journal} {Appl.
  Mech. Rev.}\ }\textbf {\bibinfo {volume} {67}},\ \bibinfo {pages} {030802}
  (\bibinfo {year} {2015})}\BibitemShut {NoStop}%
\bibitem [{\citenamefont {Seshasayanan}\ and\ \citenamefont
  {Alexakis}(2018)}]{seshasayanan2018condensates}%
  \BibitemOpen
  \bibfield  {author} {\bibinfo {author} {\bibfnamefont {K.}~\bibnamefont
  {Seshasayanan}}\ and\ \bibinfo {author} {\bibfnamefont {A.}~\bibnamefont
  {Alexakis}},\ }\href@noop {} {\bibfield  {journal} {\bibinfo  {journal} {J.
  Fluid Mech.}\ }\textbf {\bibinfo {volume} {841}},\ \bibinfo {pages} {434}
  (\bibinfo {year} {2018})}\BibitemShut {NoStop}%
\bibitem [{\citenamefont {Alexakis}(2011)}]{alexakis2011two}%
  \BibitemOpen
  \bibfield  {author} {\bibinfo {author} {\bibfnamefont {A.}~\bibnamefont
  {Alexakis}},\ }\href@noop {} {\bibfield  {journal} {\bibinfo  {journal}
  {Phys. Rev. E}\ }\textbf {\bibinfo {volume} {84}},\ \bibinfo {pages} {056330}
  (\bibinfo {year} {2011})}\BibitemShut {NoStop}%
\bibitem [{\citenamefont {Xia}\ and\ \citenamefont
  {Francois}(2017)}]{xia2017two}%
  \BibitemOpen
  \bibfield  {author} {\bibinfo {author} {\bibfnamefont {H.}~\bibnamefont
  {Xia}}\ and\ \bibinfo {author} {\bibfnamefont {N.}~\bibnamefont {Francois}},\
  }\href@noop {} {\bibfield  {journal} {\bibinfo  {journal} {Phys. Fluids}\
  }\textbf {\bibinfo {volume} {29}},\ \bibinfo {pages} {111107} (\bibinfo
  {year} {2017})}\BibitemShut {NoStop}%
\bibitem [{\citenamefont {Kellay}\ \emph {et~al.}(1995)\citenamefont {Kellay},
  \citenamefont {Wu},\ and\ \citenamefont {Goldburg}}]{kellay1995experiments}%
  \BibitemOpen
  \bibfield  {author} {\bibinfo {author} {\bibfnamefont {H.}~\bibnamefont
  {Kellay}}, \bibinfo {author} {\bibfnamefont {X.}~\bibnamefont {Wu}},\ and\
  \bibinfo {author} {\bibfnamefont {W.}~\bibnamefont {Goldburg}},\ }\href@noop
  {} {\bibfield  {journal} {\bibinfo  {journal} {Phys. Rev. Lett.}\ }\textbf
  {\bibinfo {volume} {74}},\ \bibinfo {pages} {3975} (\bibinfo {year}
  {1995})}\BibitemShut {NoStop}%
\bibitem [{\citenamefont {Smith}\ \emph {et~al.}(1996)\citenamefont {Smith},
  \citenamefont {Chasnov},\ and\ \citenamefont {Waleffe}}]{smith1996crossover}%
  \BibitemOpen
  \bibfield  {author} {\bibinfo {author} {\bibfnamefont {L.~M.}\ \bibnamefont
  {Smith}}, \bibinfo {author} {\bibfnamefont {J.~R.}\ \bibnamefont {Chasnov}},\
  and\ \bibinfo {author} {\bibfnamefont {F.}~\bibnamefont {Waleffe}},\
  }\href@noop {} {\bibfield  {journal} {\bibinfo  {journal} {Phys. Rev. Lett.}\
  }\textbf {\bibinfo {volume} {77}},\ \bibinfo {pages} {2467} (\bibinfo {year}
  {1996})}\BibitemShut {NoStop}%
\bibitem [{\citenamefont {Musacchio}\ and\ \citenamefont
  {Boffetta}(2019)}]{musacchio2019condensate}%
  \BibitemOpen
  \bibfield  {author} {\bibinfo {author} {\bibfnamefont {S.}~\bibnamefont
  {Musacchio}}\ and\ \bibinfo {author} {\bibfnamefont {G.}~\bibnamefont
  {Boffetta}},\ }\href@noop {} {\bibfield  {journal} {\bibinfo  {journal}
  {Phys. Rev. Fluids}\ }\textbf {\bibinfo {volume} {4}},\ \bibinfo {pages}
  {022602} (\bibinfo {year} {2019})}\BibitemShut {NoStop}%
\bibitem [{\citenamefont {Kolmogorov}(1941)}]{kolmogorov1941local}%
  \BibitemOpen
  \bibfield  {author} {\bibinfo {author} {\bibfnamefont {A.~N.}\ \bibnamefont
  {Kolmogorov}},\ }\href@noop {} {\bibfield  {journal} {\bibinfo  {journal} {C.
  R. Acad. Sci. URSS}\ }\textbf {\bibinfo {volume} {30}},\ \bibinfo {pages}
  {301} (\bibinfo {year} {1941})}\BibitemShut {NoStop}%
\bibitem [{\citenamefont {Frisch}(1995)}]{frisch1995}%
  \BibitemOpen
  \bibfield  {author} {\bibinfo {author} {\bibfnamefont {U.}~\bibnamefont
  {Frisch}},\ }\href@noop {} {\emph {\bibinfo {title} {Turbulence}}}\ (\bibinfo
   {publisher} {Cambridge University Press},\ \bibinfo {year}
  {1995})\BibitemShut {NoStop}%
\bibitem [{\citenamefont {Kraichnan}(1967)}]{kraichnan1967inertial}%
  \BibitemOpen
  \bibfield  {author} {\bibinfo {author} {\bibfnamefont {R.~H.}\ \bibnamefont
  {Kraichnan}},\ }\href@noop {} {\bibfield  {journal} {\bibinfo  {journal} {The
  Phys. Fluids}\ }\textbf {\bibinfo {volume} {10}},\ \bibinfo {pages} {1417}
  (\bibinfo {year} {1967})}\BibitemShut {NoStop}%
\bibitem [{\citenamefont {Batchelor}(1969)}]{batchelor1969computation}%
  \BibitemOpen
  \bibfield  {author} {\bibinfo {author} {\bibfnamefont {G.~K.}\ \bibnamefont
  {Batchelor}},\ }\href@noop {} {\bibfield  {journal} {\bibinfo  {journal}
  {Phys. Fluids}\ }\textbf {\bibinfo {volume} {12}},\ \bibinfo {pages} {II}
  (\bibinfo {year} {1969})}\BibitemShut {NoStop}%
\bibitem [{\citenamefont {Smith}\ and\ \citenamefont
  {Yakhot}(1994)}]{smith1994finite}%
  \BibitemOpen
  \bibfield  {author} {\bibinfo {author} {\bibfnamefont {L.~M.}\ \bibnamefont
  {Smith}}\ and\ \bibinfo {author} {\bibfnamefont {V.}~\bibnamefont {Yakhot}},\
  }\href@noop {} {\bibfield  {journal} {\bibinfo  {journal} {J. Fluid Mech.}\
  }\textbf {\bibinfo {volume} {274}},\ \bibinfo {pages} {115} (\bibinfo {year}
  {1994})}\BibitemShut {NoStop}%
\bibitem [{\citenamefont {Chertkov}\ \emph {et~al.}(2007)\citenamefont
  {Chertkov}, \citenamefont {Connaughton}, \citenamefont {Kolokolov},\ and\
  \citenamefont {Lebedev}}]{chertkov2007dynamics}%
  \BibitemOpen
  \bibfield  {author} {\bibinfo {author} {\bibfnamefont {M.}~\bibnamefont
  {Chertkov}}, \bibinfo {author} {\bibfnamefont {C.}~\bibnamefont
  {Connaughton}}, \bibinfo {author} {\bibfnamefont {I.}~\bibnamefont
  {Kolokolov}},\ and\ \bibinfo {author} {\bibfnamefont {V.}~\bibnamefont
  {Lebedev}},\ }\href@noop {} {\bibfield  {journal} {\bibinfo  {journal} {Phys.
  Rev. Lett.}\ }\textbf {\bibinfo {volume} {99}},\ \bibinfo {pages} {084501}
  (\bibinfo {year} {2007})}\BibitemShut {NoStop}%
\bibitem [{\citenamefont {Frishman}\ and\ \citenamefont
  {Herbert}(2018)}]{frishman2018turbulence}%
  \BibitemOpen
  \bibfield  {author} {\bibinfo {author} {\bibfnamefont {A.}~\bibnamefont
  {Frishman}}\ and\ \bibinfo {author} {\bibfnamefont {C.}~\bibnamefont
  {Herbert}},\ }\href@noop {} {\bibfield  {journal} {\bibinfo  {journal} {Phys.
  Rev. Lett.}\ }\textbf {\bibinfo {volume} {120}},\ \bibinfo {pages} {204505}
  (\bibinfo {year} {2018})}\BibitemShut {NoStop}%
\bibitem [{\citenamefont {Boffetta}\ and\ \citenamefont
  {Ecke}(2012)}]{boffetta2012two}%
  \BibitemOpen
  \bibfield  {author} {\bibinfo {author} {\bibfnamefont {G.}~\bibnamefont
  {Boffetta}}\ and\ \bibinfo {author} {\bibfnamefont {R.~E.}\ \bibnamefont
  {Ecke}},\ }\href@noop {} {\bibfield  {journal} {\bibinfo  {journal} {Annu.
  Rev. Fluid Mech.}\ }\textbf {\bibinfo {volume} {44}},\ \bibinfo {pages} {427}
  (\bibinfo {year} {2012})}\BibitemShut {NoStop}%
\bibitem [{\citenamefont {van Kan}\ and\ \citenamefont
  {Alexakis}(2019)}]{van2019condensates}%
  \BibitemOpen
  \bibfield  {author} {\bibinfo {author} {\bibfnamefont {A.}~\bibnamefont {van
  Kan}}\ and\ \bibinfo {author} {\bibfnamefont {A.}~\bibnamefont {Alexakis}},\
  }\href@noop {} {\bibfield  {journal} {\bibinfo  {journal} {J. Fluid Mech.}\
  }\textbf {\bibinfo {volume} {864}},\ \bibinfo {pages} {490} (\bibinfo {year}
  {2019})}\BibitemShut {NoStop}%
\bibitem [{\citenamefont {Vallis}\ and\ \citenamefont
  {Maltrud}(1993)}]{vallis1993generation}%
  \BibitemOpen
  \bibfield  {author} {\bibinfo {author} {\bibfnamefont {G.~K.}\ \bibnamefont
  {Vallis}}\ and\ \bibinfo {author} {\bibfnamefont {M.~E.}\ \bibnamefont
  {Maltrud}},\ }\href@noop {} {\bibfield  {journal} {\bibinfo  {journal} {J.
  Phys. Oceanogr.}\ }\textbf {\bibinfo {volume} {23}},\ \bibinfo {pages} {1346}
  (\bibinfo {year} {1993})}\BibitemShut {NoStop}%
\bibitem [{\citenamefont {Alexakis}\ and\ \citenamefont
  {Biferale}(2018)}]{alexakis2018cascades}%
  \BibitemOpen
  \bibfield  {author} {\bibinfo {author} {\bibfnamefont {A.}~\bibnamefont
  {Alexakis}}\ and\ \bibinfo {author} {\bibfnamefont {L.}~\bibnamefont
  {Biferale}},\ }\href@noop {} {\bibfield  {journal} {\bibinfo  {journal}
  {Phys. Rep.}\ } (\bibinfo {year} {2018})}\BibitemShut {NoStop}%
\bibitem [{\citenamefont {Galperin}\ \emph {et~al.}(2004)\citenamefont
  {Galperin}, \citenamefont {Nakano}, \citenamefont {Huang},\ and\
  \citenamefont {Sukoriansky}}]{galperin2004ubiquitous}%
  \BibitemOpen
  \bibfield  {author} {\bibinfo {author} {\bibfnamefont {B.}~\bibnamefont
  {Galperin}}, \bibinfo {author} {\bibfnamefont {H.}~\bibnamefont {Nakano}},
  \bibinfo {author} {\bibfnamefont {H.-P.}\ \bibnamefont {Huang}},\ and\
  \bibinfo {author} {\bibfnamefont {S.}~\bibnamefont {Sukoriansky}},\
  }\href@noop {} {\bibfield  {journal} {\bibinfo  {journal} {Geophys. Res.
  Lett.}\ }\textbf {\bibinfo {volume} {31}} (\bibinfo {year}
  {2004})}\BibitemShut {NoStop}%
\bibitem [{\citenamefont {Aurnou}\ \emph {et~al.}(2015)\citenamefont {Aurnou},
  \citenamefont {Calkins}, \citenamefont {Cheng}, \citenamefont {Julien},
  \citenamefont {King}, \citenamefont {Nieves}, \citenamefont {Soderlund},\
  and\ \citenamefont {Stellmach}}]{aurnou2015rotating}%
  \BibitemOpen
  \bibfield  {author} {\bibinfo {author} {\bibfnamefont {J.~M.}\ \bibnamefont
  {Aurnou}}, \bibinfo {author} {\bibfnamefont {M.~A.}\ \bibnamefont {Calkins}},
  \bibinfo {author} {\bibfnamefont {J.~S.}\ \bibnamefont {Cheng}}, \bibinfo
  {author} {\bibfnamefont {K.}~\bibnamefont {Julien}}, \bibinfo {author}
  {\bibfnamefont {E.~M.}\ \bibnamefont {King}}, \bibinfo {author}
  {\bibfnamefont {D.}~\bibnamefont {Nieves}}, \bibinfo {author} {\bibfnamefont
  {K.~M.}\ \bibnamefont {Soderlund}},\ and\ \bibinfo {author} {\bibfnamefont
  {S.}~\bibnamefont {Stellmach}},\ }\href@noop {} {\bibfield  {journal}
  {\bibinfo  {journal} {Phys. Earth Planet. Inter.}\ }\textbf {\bibinfo
  {volume} {246}},\ \bibinfo {pages} {52} (\bibinfo {year} {2015})}\BibitemShut
  {NoStop}%
\bibitem [{\citenamefont {Vallis}(2017)}]{vallis2017atmospheric}%
  \BibitemOpen
  \bibfield  {author} {\bibinfo {author} {\bibfnamefont {G.~K.}\ \bibnamefont
  {Vallis}},\ }\href@noop {} {\emph {\bibinfo {title} {Atmospheric and Oceanic
  Fluid Dynamics}}}\ (\bibinfo  {publisher} {Cambridge University Press},\
  \bibinfo {year} {2017})\BibitemShut {NoStop}%
\bibitem [{\citenamefont {Julien}\ \emph {et~al.}(2012)\citenamefont {Julien},
  \citenamefont {Rubio}, \citenamefont {Grooms},\ and\ \citenamefont
  {Knobloch}}]{julien2012statistical}%
  \BibitemOpen
  \bibfield  {author} {\bibinfo {author} {\bibfnamefont {K.}~\bibnamefont
  {Julien}}, \bibinfo {author} {\bibfnamefont {A.~M.}\ \bibnamefont {Rubio}},
  \bibinfo {author} {\bibfnamefont {I.}~\bibnamefont {Grooms}},\ and\ \bibinfo
  {author} {\bibfnamefont {E.}~\bibnamefont {Knobloch}},\ }\href@noop {}
  {\bibfield  {journal} {\bibinfo  {journal} {Geophys. Astrophys. Fluid Dyn.}\
  }\textbf {\bibinfo {volume} {106}},\ \bibinfo {pages} {392} (\bibinfo {year}
  {2012})}\BibitemShut {NoStop}%
\bibitem [{\citenamefont {Rubio}\ \emph {et~al.}(2014)\citenamefont {Rubio},
  \citenamefont {Julien}, \citenamefont {Knobloch},\ and\ \citenamefont
  {Weiss}}]{rubio2014upscale}%
  \BibitemOpen
  \bibfield  {author} {\bibinfo {author} {\bibfnamefont {A.~M.}\ \bibnamefont
  {Rubio}}, \bibinfo {author} {\bibfnamefont {K.}~\bibnamefont {Julien}},
  \bibinfo {author} {\bibfnamefont {E.}~\bibnamefont {Knobloch}},\ and\
  \bibinfo {author} {\bibfnamefont {J.~B.}\ \bibnamefont {Weiss}},\ }\href@noop
  {} {\bibfield  {journal} {\bibinfo  {journal} {Phys. Rev. Lett.}\ }\textbf
  {\bibinfo {volume} {112}},\ \bibinfo {pages} {144501} (\bibinfo {year}
  {2014})}\BibitemShut {NoStop}%
\bibitem [{\citenamefont {Favier}\ \emph {et~al.}(2014)\citenamefont {Favier},
  \citenamefont {Silvers},\ and\ \citenamefont {Proctor}}]{favier2014inverse}%
  \BibitemOpen
  \bibfield  {author} {\bibinfo {author} {\bibfnamefont {B.}~\bibnamefont
  {Favier}}, \bibinfo {author} {\bibfnamefont {L.~J.}\ \bibnamefont
  {Silvers}},\ and\ \bibinfo {author} {\bibfnamefont {M.~R.~E.}\ \bibnamefont
  {Proctor}},\ }\href@noop {} {\bibfield  {journal} {\bibinfo  {journal} {Phys.
  Fluids}\ }\textbf {\bibinfo {volume} {26}},\ \bibinfo {pages} {096605}
  (\bibinfo {year} {2014})}\BibitemShut {NoStop}%
\bibitem [{\citenamefont {Guervilly}\ \emph {et~al.}(2014)\citenamefont
  {Guervilly}, \citenamefont {Hughes},\ and\ \citenamefont
  {Jones}}]{guervilly2014large}%
  \BibitemOpen
  \bibfield  {author} {\bibinfo {author} {\bibfnamefont {C.}~\bibnamefont
  {Guervilly}}, \bibinfo {author} {\bibfnamefont {D.~W.}\ \bibnamefont
  {Hughes}},\ and\ \bibinfo {author} {\bibfnamefont {C.~A.}\ \bibnamefont
  {Jones}},\ }\href@noop {} {\bibfield  {journal} {\bibinfo  {journal} {J.
  Fluid Mech.}\ }\textbf {\bibinfo {volume} {758}},\ \bibinfo {pages} {407}
  (\bibinfo {year} {2014})}\BibitemShut {NoStop}%
\bibitem [{\citenamefont {Stellmach}\ \emph {et~al.}(2014)\citenamefont
  {Stellmach}, \citenamefont {Lischper}, \citenamefont {Julien}, \citenamefont
  {Vasil}, \citenamefont {Cheng}, \citenamefont {Ribeiro}, \citenamefont
  {King},\ and\ \citenamefont {Aurnou}}]{stellmach2014approaching}%
  \BibitemOpen
  \bibfield  {author} {\bibinfo {author} {\bibfnamefont {S.}~\bibnamefont
  {Stellmach}}, \bibinfo {author} {\bibfnamefont {M.}~\bibnamefont {Lischper}},
  \bibinfo {author} {\bibfnamefont {K.}~\bibnamefont {Julien}}, \bibinfo
  {author} {\bibfnamefont {G.}~\bibnamefont {Vasil}}, \bibinfo {author}
  {\bibfnamefont {J.~S.}\ \bibnamefont {Cheng}}, \bibinfo {author}
  {\bibfnamefont {A.}~\bibnamefont {Ribeiro}}, \bibinfo {author} {\bibfnamefont
  {E.~M.}\ \bibnamefont {King}},\ and\ \bibinfo {author} {\bibfnamefont
  {J.~M.}\ \bibnamefont {Aurnou}},\ }\href@noop {} {\bibfield  {journal}
  {\bibinfo  {journal} {Phys. Rev. Lett.}\ }\textbf {\bibinfo {volume} {113}},\
  \bibinfo {pages} {254501} (\bibinfo {year} {2014})}\BibitemShut {NoStop}%
\bibitem [{\citenamefont {Kunnen}\ \emph {et~al.}(2016)\citenamefont {Kunnen},
  \citenamefont {Ostilla-M{\'o}nico}, \citenamefont {Van Der~Poel},
  \citenamefont {Verzicco},\ and\ \citenamefont
  {Lohse}}]{kunnen2016transition}%
  \BibitemOpen
  \bibfield  {author} {\bibinfo {author} {\bibfnamefont {R.~P.~J.}\
  \bibnamefont {Kunnen}}, \bibinfo {author} {\bibfnamefont {R.}~\bibnamefont
  {Ostilla-M{\'o}nico}}, \bibinfo {author} {\bibfnamefont {E.~P.}\ \bibnamefont
  {Van Der~Poel}}, \bibinfo {author} {\bibfnamefont {R.}~\bibnamefont
  {Verzicco}},\ and\ \bibinfo {author} {\bibfnamefont {D.}~\bibnamefont
  {Lohse}},\ }\href@noop {} {\bibfield  {journal} {\bibinfo  {journal} {J.
  Fluid Mech.}\ }\textbf {\bibinfo {volume} {799}},\ \bibinfo {pages} {413}
  (\bibinfo {year} {2016})}\BibitemShut {NoStop}%
\bibitem [{\citenamefont {Julien}\ \emph {et~al.}(2016)\citenamefont {Julien},
  \citenamefont {Aurnou}, \citenamefont {Calkins}, \citenamefont {Knobloch},
  \citenamefont {Marti}, \citenamefont {Stellmach},\ and\ \citenamefont
  {Vasil}}]{julien2016nonlinear}%
  \BibitemOpen
  \bibfield  {author} {\bibinfo {author} {\bibfnamefont {K.}~\bibnamefont
  {Julien}}, \bibinfo {author} {\bibfnamefont {J.~M.}\ \bibnamefont {Aurnou}},
  \bibinfo {author} {\bibfnamefont {M.~A.}\ \bibnamefont {Calkins}}, \bibinfo
  {author} {\bibfnamefont {E.}~\bibnamefont {Knobloch}}, \bibinfo {author}
  {\bibfnamefont {P.}~\bibnamefont {Marti}}, \bibinfo {author} {\bibfnamefont
  {S.}~\bibnamefont {Stellmach}},\ and\ \bibinfo {author} {\bibfnamefont
  {G.~M.}\ \bibnamefont {Vasil}},\ }\href@noop {} {\bibfield  {journal}
  {\bibinfo  {journal} {J. Fluid Mech.}\ }\textbf {\bibinfo {volume} {798}},\
  \bibinfo {pages} {50} (\bibinfo {year} {2016})}\BibitemShut {NoStop}%
\bibitem [{\citenamefont {Plumley}\ \emph {et~al.}(2016)\citenamefont
  {Plumley}, \citenamefont {Julien}, \citenamefont {Marti},\ and\ \citenamefont
  {Stellmach}}]{plumley2016effects}%
  \BibitemOpen
  \bibfield  {author} {\bibinfo {author} {\bibfnamefont {M.}~\bibnamefont
  {Plumley}}, \bibinfo {author} {\bibfnamefont {K.}~\bibnamefont {Julien}},
  \bibinfo {author} {\bibfnamefont {P.}~\bibnamefont {Marti}},\ and\ \bibinfo
  {author} {\bibfnamefont {S.}~\bibnamefont {Stellmach}},\ }\href@noop {}
  {\bibfield  {journal} {\bibinfo  {journal} {J. Fluid Mech.}\ }\textbf
  {\bibinfo {volume} {803}},\ \bibinfo {pages} {51} (\bibinfo {year}
  {2016})}\BibitemShut {NoStop}%
\bibitem [{Note1()}]{Note1}%
  \BibitemOpen
  \bibinfo {note} {The slight difference in $Pr$ between the two simulation
  series is for comparison with (ongoing) experiments in our group \cite
  {cheng2018heuristic}.}\BibitemShut {Stop}%
\bibitem [{\citenamefont {Ostilla-Monico}\ \emph {et~al.}(2015)\citenamefont
  {Ostilla-Monico}, \citenamefont {Yang}, \citenamefont {van~der Poel},
  \citenamefont {Lohse},\ and\ \citenamefont {Verzicco}}]{ostilla2015multiple}%
  \BibitemOpen
  \bibfield  {author} {\bibinfo {author} {\bibfnamefont {R.}~\bibnamefont
  {Ostilla-Monico}}, \bibinfo {author} {\bibfnamefont {Y.}~\bibnamefont
  {Yang}}, \bibinfo {author} {\bibfnamefont {E.~P.}\ \bibnamefont {van~der
  Poel}}, \bibinfo {author} {\bibfnamefont {D.}~\bibnamefont {Lohse}},\ and\
  \bibinfo {author} {\bibfnamefont {R.}~\bibnamefont {Verzicco}},\ }\href@noop
  {} {\bibfield  {journal} {\bibinfo  {journal} {J. Comput. Phys.}\ }\textbf
  {\bibinfo {volume} {301}},\ \bibinfo {pages} {308} (\bibinfo {year}
  {2015})}\BibitemShut {NoStop}%
\bibitem [{\citenamefont
  {Chandrasekhar}(1961)}]{chandrasekhar2013hydrodynamic}%
  \BibitemOpen
  \bibfield  {author} {\bibinfo {author} {\bibfnamefont {S.}~\bibnamefont
  {Chandrasekhar}},\ }\href@noop {} {\emph {\bibinfo {title} {Hydrodynamic and
  Hydromagnetic Stability}}}\ (\bibinfo  {publisher} {Oxford University
  Press},\ \bibinfo {year} {1961})\BibitemShut {NoStop}%
\bibitem [{Note2()}]{Note2}%
  \BibitemOpen
  \bibinfo {note} {The critical Rayleigh number for onset of oscillatory
  convection ($Pr<0.68$) is $Ra_c=17.4(Ek/Pr)^{-4/3}$; for onset of steady
  convection ($Pr\ge 0.68$) it is $Ra_c=8.7Ek^{-4/3}$ \cite
  {chandrasekhar2013hydrodynamic}.}\BibitemShut {Stop}%
\bibitem [{\citenamefont {Cheng}\ \emph {et~al.}(2015)\citenamefont {Cheng},
  \citenamefont {Stellmach}, \citenamefont {Ribeiro}, \citenamefont {Grannan},
  \citenamefont {King},\ and\ \citenamefont {Aurnou}}]{cheng2015laboratory}%
  \BibitemOpen
  \bibfield  {author} {\bibinfo {author} {\bibfnamefont {J.~S.}\ \bibnamefont
  {Cheng}}, \bibinfo {author} {\bibfnamefont {S.}~\bibnamefont {Stellmach}},
  \bibinfo {author} {\bibfnamefont {A.}~\bibnamefont {Ribeiro}}, \bibinfo
  {author} {\bibfnamefont {A.}~\bibnamefont {Grannan}}, \bibinfo {author}
  {\bibfnamefont {E.~M.}\ \bibnamefont {King}},\ and\ \bibinfo {author}
  {\bibfnamefont {J.~M.}\ \bibnamefont {Aurnou}},\ }\href@noop {} {\bibfield
  {journal} {\bibinfo  {journal} {Geophys. J. Int.}\ }\textbf {\bibinfo
  {volume} {201}},\ \bibinfo {pages} {1} (\bibinfo {year} {2015})}\BibitemShut
  {NoStop}%
\bibitem [{\citenamefont {Nieves}\ \emph {et~al.}(2014)\citenamefont {Nieves},
  \citenamefont {Rubio},\ and\ \citenamefont {Julien}}]{nieves2014statistical}%
  \BibitemOpen
  \bibfield  {author} {\bibinfo {author} {\bibfnamefont {D.}~\bibnamefont
  {Nieves}}, \bibinfo {author} {\bibfnamefont {A.~M.}\ \bibnamefont {Rubio}},\
  and\ \bibinfo {author} {\bibfnamefont {K.}~\bibnamefont {Julien}},\
  }\href@noop {} {\bibfield  {journal} {\bibinfo  {journal} {Phys. Fluids}\
  }\textbf {\bibinfo {volume} {26}},\ \bibinfo {pages} {086602} (\bibinfo
  {year} {2014})}\BibitemShut {NoStop}%
\bibitem [{\citenamefont {Kunnen}\ \emph
  {et~al.}(2010{\natexlab{a}})\citenamefont {Kunnen}, \citenamefont {Geurts},\
  and\ \citenamefont {Clercx}}]{kunnen2010experimental}%
  \BibitemOpen
  \bibfield  {author} {\bibinfo {author} {\bibfnamefont {R.~P.~J.}\
  \bibnamefont {Kunnen}}, \bibinfo {author} {\bibfnamefont {B.~J.}\
  \bibnamefont {Geurts}},\ and\ \bibinfo {author} {\bibfnamefont {H.~J.~H.}\
  \bibnamefont {Clercx}},\ }\href@noop {} {\bibfield  {journal} {\bibinfo
  {journal} {J. Fluid Mech.}\ }\textbf {\bibinfo {volume} {642}},\ \bibinfo
  {pages} {445} (\bibinfo {year} {2010}{\natexlab{a}})}\BibitemShut {NoStop}%
\bibitem [{\citenamefont {Kunnen}\ \emph
  {et~al.}(2010{\natexlab{b}})\citenamefont {Kunnen}, \citenamefont {Clercx},\
  and\ \citenamefont {Geurts}}]{kunnen2010vortex}%
  \BibitemOpen
  \bibfield  {author} {\bibinfo {author} {\bibfnamefont {R.~P.~J.}\
  \bibnamefont {Kunnen}}, \bibinfo {author} {\bibfnamefont {H.~J.~H.}\
  \bibnamefont {Clercx}},\ and\ \bibinfo {author} {\bibfnamefont {B.~J.}\
  \bibnamefont {Geurts}},\ }\href@noop {} {\bibfield  {journal} {\bibinfo
  {journal} {Phys. Rev. E}\ }\textbf {\bibinfo {volume} {82}},\ \bibinfo
  {pages} {036306} (\bibinfo {year} {2010}{\natexlab{b}})}\BibitemShut
  {NoStop}%
\bibitem [{\citenamefont {Hopfinger}\ \emph {et~al.}(1982)\citenamefont
  {Hopfinger}, \citenamefont {Browand},\ and\ \citenamefont
  {Gagne}}]{hopfinger1982turbulence}%
  \BibitemOpen
  \bibfield  {author} {\bibinfo {author} {\bibfnamefont {E.~J.}\ \bibnamefont
  {Hopfinger}}, \bibinfo {author} {\bibfnamefont {F.~K.}\ \bibnamefont
  {Browand}},\ and\ \bibinfo {author} {\bibfnamefont {Y.}~\bibnamefont
  {Gagne}},\ }\href@noop {} {\bibfield  {journal} {\bibinfo  {journal} {J.
  Fluid Mech.}\ }\textbf {\bibinfo {volume} {125}},\ \bibinfo {pages} {505}
  (\bibinfo {year} {1982})}\BibitemShut {NoStop}%
\bibitem [{\citenamefont {Baroud}\ \emph {et~al.}(2003)\citenamefont {Baroud},
  \citenamefont {Plapp}, \citenamefont {Swinney},\ and\ \citenamefont
  {She}}]{baroud2003scaling}%
  \BibitemOpen
  \bibfield  {author} {\bibinfo {author} {\bibfnamefont {C.~N.}\ \bibnamefont
  {Baroud}}, \bibinfo {author} {\bibfnamefont {B.~B.}\ \bibnamefont {Plapp}},
  \bibinfo {author} {\bibfnamefont {H.~L.}\ \bibnamefont {Swinney}},\ and\
  \bibinfo {author} {\bibfnamefont {Z.-S.}\ \bibnamefont {She}},\ }\href@noop
  {} {\bibfield  {journal} {\bibinfo  {journal} {Phys. Fluids}\ }\textbf
  {\bibinfo {volume} {15}},\ \bibinfo {pages} {2091} (\bibinfo {year}
  {2003})}\BibitemShut {NoStop}%
\bibitem [{\citenamefont {Sreenivasan}\ and\ \citenamefont
  {Davidson}(2008)}]{sreenivasan2008formation}%
  \BibitemOpen
  \bibfield  {author} {\bibinfo {author} {\bibfnamefont {B.}~\bibnamefont
  {Sreenivasan}}\ and\ \bibinfo {author} {\bibfnamefont {P.~A.}\ \bibnamefont
  {Davidson}},\ }\href@noop {} {\bibfield  {journal} {\bibinfo  {journal}
  {Phys. Fluids}\ }\textbf {\bibinfo {volume} {20}},\ \bibinfo {pages} {085104}
  (\bibinfo {year} {2008})}\BibitemShut {NoStop}%
\bibitem [{\citenamefont {Alexakis}\ \emph {et~al.}(2005)\citenamefont
  {Alexakis}, \citenamefont {Mininni},\ and\ \citenamefont
  {Pouquet}}]{alexakis2005shell}%
  \BibitemOpen
  \bibfield  {author} {\bibinfo {author} {\bibfnamefont {A.}~\bibnamefont
  {Alexakis}}, \bibinfo {author} {\bibfnamefont {P.~D.}\ \bibnamefont
  {Mininni}},\ and\ \bibinfo {author} {\bibfnamefont {A.}~\bibnamefont
  {Pouquet}},\ }\href@noop {} {\bibfield  {journal} {\bibinfo  {journal} {Phys.
  Rev. E}\ }\textbf {\bibinfo {volume} {72}},\ \bibinfo {pages} {046301}
  (\bibinfo {year} {2005})}\BibitemShut {NoStop}%
\bibitem [{\citenamefont {Mininni}\ \emph {et~al.}(2009)\citenamefont
  {Mininni}, \citenamefont {Alexakis},\ and\ \citenamefont
  {Pouquet}}]{mininni2009scale}%
  \BibitemOpen
  \bibfield  {author} {\bibinfo {author} {\bibfnamefont {P.~D.}\ \bibnamefont
  {Mininni}}, \bibinfo {author} {\bibfnamefont {A.}~\bibnamefont {Alexakis}},\
  and\ \bibinfo {author} {\bibfnamefont {A.}~\bibnamefont {Pouquet}},\
  }\href@noop {} {\bibfield  {journal} {\bibinfo  {journal} {Phys. Fluids}\
  }\textbf {\bibinfo {volume} {21}},\ \bibinfo {pages} {015108} (\bibinfo
  {year} {2009})}\BibitemShut {NoStop}%
\bibitem [{\citenamefont {Deardorff}\ and\ \citenamefont
  {Willis}(1967)}]{deardorff1967investigation}%
  \BibitemOpen
  \bibfield  {author} {\bibinfo {author} {\bibfnamefont {J.~W.}\ \bibnamefont
  {Deardorff}}\ and\ \bibinfo {author} {\bibfnamefont {G.~E.}\ \bibnamefont
  {Willis}},\ }\href@noop {} {\bibfield  {journal} {\bibinfo  {journal} {J.
  Fluid Mech.}\ }\textbf {\bibinfo {volume} {28}},\ \bibinfo {pages} {675}
  (\bibinfo {year} {1967})}\BibitemShut {NoStop}%
\bibitem [{\citenamefont {Kerr}(2001)}]{kerr2001energy}%
  \BibitemOpen
  \bibfield  {author} {\bibinfo {author} {\bibfnamefont {R.~M.}\ \bibnamefont
  {Kerr}},\ }\href@noop {} {\bibfield  {journal} {\bibinfo  {journal} {Phys.
  Rev. Lett.}\ }\textbf {\bibinfo {volume} {87}},\ \bibinfo {pages} {244502}
  (\bibinfo {year} {2001})}\BibitemShut {NoStop}%
\bibitem [{\citenamefont {Kunnen}\ \emph {et~al.}(2009)\citenamefont {Kunnen},
  \citenamefont {Geurts},\ and\ \citenamefont {Clercx}}]{kunnen2009turbulence}%
  \BibitemOpen
  \bibfield  {author} {\bibinfo {author} {\bibfnamefont {R.~P.~J.}\
  \bibnamefont {Kunnen}}, \bibinfo {author} {\bibfnamefont {B.~J.}\
  \bibnamefont {Geurts}},\ and\ \bibinfo {author} {\bibfnamefont {H.~J.~H.}\
  \bibnamefont {Clercx}},\ }\href@noop {} {\bibfield  {journal} {\bibinfo
  {journal} {Eur. J. Mech. B/Fluids}\ }\textbf {\bibinfo {volume} {28}},\
  \bibinfo {pages} {578} (\bibinfo {year} {2009})}\BibitemShut {NoStop}%
\bibitem [{\citenamefont {Pope}(2000)}]{pope2001turbulent}%
  \BibitemOpen
  \bibfield  {author} {\bibinfo {author} {\bibfnamefont {S.~B.}\ \bibnamefont
  {Pope}},\ }\href@noop {} {\bibinfo {title} {Turbulent flows}} (\bibinfo
  {year} {2000})\BibitemShut {NoStop}%
\bibitem [{\citenamefont {Xia}\ \emph {et~al.}(2011)\citenamefont {Xia},
  \citenamefont {Byrne}, \citenamefont {Falkovich},\ and\ \citenamefont
  {Shats}}]{xia2011upscale}%
  \BibitemOpen
  \bibfield  {author} {\bibinfo {author} {\bibfnamefont {H.}~\bibnamefont
  {Xia}}, \bibinfo {author} {\bibfnamefont {D.}~\bibnamefont {Byrne}}, \bibinfo
  {author} {\bibfnamefont {G.}~\bibnamefont {Falkovich}},\ and\ \bibinfo
  {author} {\bibfnamefont {M.}~\bibnamefont {Shats}},\ }\href@noop {}
  {\bibfield  {journal} {\bibinfo  {journal} {Nat. Phys.}\ }\textbf {\bibinfo
  {volume} {7}},\ \bibinfo {pages} {321} (\bibinfo {year} {2011})}\BibitemShut
  {NoStop}%
\bibitem [{\citenamefont {Cheng}\ \emph {et~al.}(2018)\citenamefont {Cheng},
  \citenamefont {Aurnou}, \citenamefont {Julien},\ and\ \citenamefont
  {Kunnen}}]{cheng2018heuristic}%
  \BibitemOpen
  \bibfield  {author} {\bibinfo {author} {\bibfnamefont {J.~S.}\ \bibnamefont
  {Cheng}}, \bibinfo {author} {\bibfnamefont {J.~M.}\ \bibnamefont {Aurnou}},
  \bibinfo {author} {\bibfnamefont {K.}~\bibnamefont {Julien}},\ and\ \bibinfo
  {author} {\bibfnamefont {R.~P.~J.}\ \bibnamefont {Kunnen}},\ }\href@noop {}
  {\bibfield  {journal} {\bibinfo  {journal} {Geophys. Astrophys. Fluid Dyn.}\
  }\textbf {\bibinfo {volume} {112}},\ \bibinfo {pages} {277} (\bibinfo {year}
  {2018})}\BibitemShut {NoStop}%
\end{thebibliography}
\end{document}